\documentclass[fleqn,usenatbib]{mnras}

\usepackage{newtxtext,newtxmath,widetext,flushend}
\usepackage[T1]{fontenc}
\usepackage{ae,aecompl}

\usepackage{graphicx}	% Including figure files
\usepackage{amsmath}	% Advanced maths commands
\usepackage{amssymb}	% Extra maths symbols

\usepackage{epsfig}
\usepackage{latexsym}

\def\beq{\begin{equation}}
\def\eeq{\end{equation}}
\def\br{\begin{eqnarray}}
\def\er{\end{eqnarray}}
\def\benu{\begin{enumerate}}
\def\efnu{\end{enumerate}}
\def\nn{\nonumber}

\def\l{\left}
\def\r{\right}

\def\d{{\rm d}}

\def \lleq {\lower0.9ex\hbox{ $\buildrel < \over \sim$} ~}
\def \ggeq {\lower0.9ex\hbox{ $\buildrel > \over \sim$} ~}

\title[Metastable Dark Energy]{Metastable Dark Energy with Radioactive-like Decay}

\author[A. Shafieloo et al.]{
Arman Shafieloo,$^{1,2}$\thanks{E-mail: shafieloo@kasi.re.kr}
Dhiraj Kumar Hazra,$^{3}$
Varun Sahni$^{4}$
and Alexei A. Starobinsky$^{5,6}$
\\
$^{1}$Korea Astronomy and Space Science Institute, Daejeon 34055, Korea\\
$^{2}$University of Science and Technology, Daejeon 34113, Korea\\
$^{3}$AstroParticule et Cosmologie (APC)/Paris Centre for Cosmological Physics, Universit\'e
Paris Diderot, CNRS, CEA,\\ Observatoire de Paris, Sorbonne Paris Cit\'e University, 10, rue Alice Domon et Leonie Duquet, 75205 Paris Cedex 13, France\\
$^{4}$Inter-University Centre for Astronomy and Astrophysics, Post Bag 4, Ganeshkhind, Pune 411 007, India\\
$^{5}$Landau Institute for Theoretical Physics RAS, Moscow, 119334, Russia\\
$^{6}$Bogolyubov Laboratory of Theoretical Physics, Joint Institute for Nuclear Research, Dubna 141980, Russia
}

\date{Accepted XXX. Received YYY; in original form ZZZ}

\pubyear{2017}

\begin{document}
\label{firstpage}
\pagerange{\pageref{firstpage}--\pageref{lastpage}}
\maketitle

% Abstract of the paper
\begin{abstract}
We propose a new class of metastable dark energy (DE) phenomenological models in which
the DE decay rate does not depend on external parameters such as the scale
factor or the curvature of the Universe. Instead, the DE decay rate is assumed to be a constant depending only on 
intrinsic properties of DE and the type of a decay channel, similar to 
case of the radioactive decay of unstable particles and nuclei.
As a consequence, the DE energy density becomes a function of the 
proper time elapsed since its formation, presumably in the very early
Universe. Such a natural type of DE decay can profoundly affect the expansion history of the 
Universe and its age. Metastable DE can decay in three distinct ways: (i) exponentially, (ii) into dark matter, (iii) into dark radiation.
Testing metastable DE models with observational data we find that the decay half-life must be many times larger
 than the age of the Universe. 
Models in which dark energy decays into dark matter lead to lower values of 
the Hubble parameter at large redshifts relative to $\Lambda$CDM. Consequently
 these models provide a better fit to cosmological BAO
data (especially data from from high redshift quasars) than concordance ($\Lambda$CDM) cosmology.
\end{abstract}

% Select between one and six entries from the list of approved keywords.
% Don't make up new ones.

%%%%%%%%%%%%%%%%%%%%%%%%%%%%%%%%%%%%%%%%%%%%%%%%%%

%%%%%%%%%%%%%%%%% BODY OF PAPER %%%%%%%%%%%%%%%%%%
\section{Introduction}

One of the main goals of physical cosmology is to understand the nature of
the constituents of our Universe.
Of these perhaps the most enigmatic are dark matter and dark energy (DE)
which, together, constitute nearly $96\%$ 
of the total density of the Universe \citep{Sahni:1999gb,Sahni:2004ai}. 
While many theoretical models have been advanced to explain the nature of the dark
Universe, what has firmly been established 
is that dark matter clusters and has a pressureless equation of state,
while dark energy possesses negative pressure which can cause the Universe to accelerate
\citep{Carroll:2000fy,Peebles:2002gy,Padmanabhan:2002ji,Sahni:2005pf,Copeland:2006wr,Sahni:2006pa}. It is also widely believed that both dark matter and dark energy have a non-baryonic origin.
%In the absence of a firm theoretical underpinning for dark matter and dark energy
The current debate on the nature of the dark Universe allows for the fact
%it has been suggested 
that the two dark components might interact with each other. 
The transfer of energy between dark energy and dark matter could lead to
interesting, and possibly unique, observational signatures; see \citep{Amendola:1999er,Guo:2007zk,Boehmer:2008av,Valiviita:2008iv,He:2008tn,Micheletti:2009pk,He:2010im,Pavan:2011xn,Faraoni:2014vra,Salvatelli:2014zta,Abdalla:2014cla} and references therein.

Existing observational data do not show any statistically significant
deviation from concordance cosmology in which DE coincides with an exact 
cosmological constant $\Lambda$ and is therefore stable and eternal. 
However, the remarkable {\em{qualitative}} similarity between the physical
properties of current DE and the primordial DE that drove inflation in the 
very early Universe (the latter without a doubt being metastable) 
makes it rather natural (though not
obligatory) to put forward the hypothesis that the present DE is
metastable and not eternal, too \footnote{Here we call DE {\em any} driving force for an accelerated expansion of the Universe, irrespective of its nature. Using the terminology introduced in \citep{Sahni:2006pa}, DE can be physical (a new field of matter) or geometrical (modified gravity), but most generally it is mixed one like in the case of scalar-tensor gravity.}. Historically, such a hypothesis was
first advanced in a very old paper by the Soviet physicist Matvei
Bronstein in 1933 \citep{Bronstein:1933} \footnote{Since this was shortly before
the understanding of the structure of atomic nuclei and the existence of
strong interactions, Bronstein tried to use the instability of $\Lambda$ 
as a source for the energy of starlight. Still his estimate of the decay 
rate, considered simply as a (unjustified) hypothesis, was
$\dot\Lambda/\Lambda \sim 10^{-24}$ s$^{-1}$, which is so much
smaller than the inverse of the present age of the Universe that it cannot
be excluded by existing observational data.}. More realistic models of
DE which differed from an exact $\Lambda$-term began to appear in the 1980's 
and there are plenty of them by now.

However, in most of these models beginning from \citep{Ozer:1985ws}, the DE 
energy density was assumed to be determined by physical properties 
external to DE itself, for instance DE could depend upon the scale factor of the
Friedmann-Lemaitre-Robertson-Walker (FLRW) universe $a(t)$, its
expansion rate $H(t)$, scalar curvature $R$, etc. \footnote{The 
$\Lambda(H)$ cosmological model \citep{Shapiro:2000dz} and its recent 
"decaying vacuum" extensions such as \citep{Lima:2012mu} also belong to this
class.} On the other hand, unstable nuclei and elementary
particles decay exponentially with time with decay rates not
depending on external conditions but defined by their intrinsic
composition and structure only. This simple and observed manner of
decay has attracted somewhat little attention in studies of
the possible decay of DE. In particular, one can hardly
find a lower limit on the (decay channel dependent) DE half-life
in handbooks, in contrast for instance, to the same quantity for
the proton. That is why this paper is devoted to the study of this
problem for several models (channels) of DE decay. Using this nuclear physics analogy, we shall call all channels of DE 
decay in which the decay rate is a constant not depending on space-time 
metric and curvature radioactive-like ones. We assume
that DE decays totally (or almost totally) into other dark
(i.e. not participating in strong, electromagnetic and weak
interactions) constituents, since the decay of DE into visible
matter is strongly restricted by observations. Note that our model of metastable DE closely resembles the decay of dark matter in models in which the latter is composed of metastable particles such as a sterile neutrino. It is well known that a sterile neutrino can decay through the mixing with virtual ("off mass shell") neutrinos and there are even reports that such a decay may be responsible for the emission line at 3.52 keV in the x-ray spectra of the Andromeda galaxy and the Perseus galaxy cluster \citep{Boyarsky:2014jta}.

It follows from the assumption that the DE decay rate is a constant and
depends only on its internal composition (in particular, it does not 
depend on the DE energy density $\rho_{DE}$) that $\rho_{DE}$ should be
a function of the proper time elapsed since the formation of DE (which
we assume occured in the early Universe). Here it is important to
emphasize that the energy-momentum tensor of both decaying DE and 
its decay products should be taken into account in the FLRW
equations for $a(t)$ to avoid inconsistency with the Bianchi
identities \footnote{It is interesting that a similar critical
remark had been immediately made to M. Bronstein by L. Landau, 
and this was mentioned in the 'Note added' to the paper \citep{Bronstein:1933}}.
Thus, the equation of state $w\equiv p/\rho$ of the total mixture
of DE and its decay products may not be $w_{tot}=-1$ as is the case for 
$\Lambda$. In fact, $w_{tot}>-1$. This allows one to
find a system of reference where this mixture is at rest on the average,
and the time in terms of which the DE half-life time is measured is
the proper time in this system. We assume that this system is at rest
in the FLRW frame, so this time is the usual cosmic time $t$.
\footnote{In the presence of isocurvature fluctuations, it is
possible that this system has some velocity with respect to the
FLRW one, and then part of the CMB temperature dipole is primordial.
However, in the absence of extreme fine-tuning, this velocity is 
non-relativistic and small, so we may neglect it for our purpose.}

This 'dark mixture' can be described either totally as one component,
or as a sum of two interacting components. In the first case, if some 
time dependence of $\rho_{DE}(t)$ is assumed (we keep the name and
index 'DE' for this mixture for simplicity and consider the
exponential law in the first approximation), the pressure $p_{DE}$
is defined from the conservation equation as
\begin{equation}
p_{DE}=-\frac{\dot \rho_{DE}+3H\rho_{DE}}{3H}=
-\rho_{DE}\left(1-\frac{\Gamma}{3H}\right)
\label{cons}
\end{equation}
where $H\equiv \dot a/a$ is the Hubble parameter and $\Gamma={\rm const}$ is 
the decay rate. This is our model I. In the second case, we may
put $w_1\equiv w_{DE}=-1$ in the first approximation, and the products
of its decay are modeled by either dust-like dark matter with $w_2=0$
(model II) or by dark radiation with $w_3=1/3$ (Model III). Since the aim of the paper is to confront three purely phenomenological ways of DE decay with observational data, we don't set a goal to construct microphysical models producing exactly the same behavior of $\rho_{DE}(t)$ here. Still in the case of the Model I, a similar behavior naturally occurs in the case of slow rolling quintessence, i.e. if DE is modeled by a minimally coupled scalar field with a sufficiently flat potential, and we discuss it in more details in the Appendix.

%In this paper we present a new model of dark energy in which the latter is metastable
%and can decay either exponentially (hence having an evolving effective $w(z) < -1/3$) or into dark matter
%($w = 0$) or even dark radiation ($w \simeq 1/3$).
In section~\ref{sec:formalism} we discuss these models,
while in section~\ref{sec:analysis} we compare their properties against observations
and constrain their free parameters.
Our results are presented in section~\ref{sec:results} and
conclusions are drawn in section~\ref{sec:discussion}. 
Broadly speaking our analysis shows that if dark energy is metastable then its
decay `half-life' should exceed the age of the Universe.
We find that the decay of dark energy into dark matter 
alleviates much of the tension faced by concordance cosmology ($\Lambda$CDM) when simultaneously
fitting CMB data and
BAO data from high redshift quasars.\\
 
 In the Appendix, explicit expressions for the Hubble factor behaviour as
a function of redshift are presented in the limit $\Gamma\ll H_0$. 
 
%%%%%%%%%%%%%%%%%%%%%%%%%%%%%%%%%%%%%%%%%%%%%%%%%%%%%%%%%%%%%%%%%%%%%%%%%%%%%%%

\section{Formalism}~\label{sec:formalism}

In this paper we consider three models of metastable dark energy (DE). In Model I, DE decays exponentially and hence has an 
evolving effective equation of state (EOS). In Model II, DE decays into dark matter, while in Model III, DE decays into dark 
radiation. We assume that the decay of DE is due to its {\em intrinsic properties} and is not related to the expansion of the 
Universe. In particular it could also occur in flat space-time with $H=0$. For simplicity we assume that DE has the equation 
of state $w=-1$ at high redshift, although the methods developed in our paper can easily be generalised to other DE models.

\subsection{Model I : Exponentially decaying DE}
We assume a `radioactive decay' scheme for the time-evolution of dark energy.
In other words, the present value of the DE density, $\rho_{\rm DE}(t_0)$, 
is related to its value at an earlier time, $\rho_{\rm DE}(t)$, by
\beq
\rho_{\rm DE}(t)=\rho_{\rm DE}(t_0)\times \exp[-\Gamma(t-t_0)],~\label{eq:model-I}
\eeq
here, the $\Gamma$ is the only free parameter in the equation.
Eqn (\ref{eq:model-I}) follows from 
\beq
\dot\rho_{DE}=-\Gamma\rho_{DE}.
\label{eq:2}
\eeq
From (\ref{eq:model-I}) one finds the
decay `half life' of DE to be
$t_{1/2}=\ln (2)/\Gamma$, when $\Gamma > 0$. Negative values of $\Gamma$ imply an
increase in the DE density with time which corresponds to phantom-like behaviour.
Note that $\Gamma$ has the dimensions of inverse time. 

The Hubble parameter, $h(z)=H(z)/H_0$, is easily obtained from the FRW equation 
to be
\beq
h^2(z)=(1-\Omega_{\rm 0m})\exp\l[\frac{\Gamma}{H_0}\int^{z}_{0}\frac{\d z}{(1+z) h(z)}\r]+\Omega_{\rm 0m} (1+z)^{3}.~\label{eq:rad_de_hz}
\eeq
In what follows we shall solve this equation iteratively to determine
 the expansion history of the Universe. 
Note that DE can be described by the dimensionless free parameter 
${\Gamma}/{H_0}$. Hence the decay 
rate (or growth rate) of DE can be described without knowing the value of $H_0$. 
This is significant 
since supernovae (SNe) data can directly constrain the decay constant ${\Gamma}/{H_0}$, thereby describing 
the half-life of dark energy in units of the age of the Universe.

It is seen from Eq. \eqref{cons} that the phenomenological decay law
(2.1) does not correspond to a constant equation of state $w_{DE}$.
Moreover, $w_{DE}$ exceeds unity for $H<\Gamma/6$. This shows that this 
law, if taken literally for all times, might require a rather unusual 
microscopic model. However, first, we shall use it around the present
time only, when $\Gamma < H_0$ and probably $\Gamma\ll H_0$, so that
the fraction of already decayed DE is small. Second, $w_{DE}>1$ is not
prohibited by causality. E.g., it can be easily realized by a usual
scalar field with a negative potential minimally coupled to gravity.
Thus, because of the ubiquitous appearance of this law in atomic and
particle physics, it makes sense to confront it with observations in
the case of decaying DE.

%One should note that in this model dark energy is decaying exponentially without interacting with other energy components such as matter 
%or radiation. While it may look like that we are violating the energy-momentum conservation, this is not the case as this model is algebraically 
%equivalent to an evolving dark energy model with $w(t)= \frac{\Gamma}{ 3H(t)} - 1$ which is the same as assuming $\Gamma=3[1+w(t)]H(t)$. 

%{\bf Varun: I think the above equn's are incorrect. If DE is decaying into DE with a different
%EOS, say $w$, then in keeping with the next subsection, the equations should be

%\br
%\dot\rho^{(1)}_{DE}=-\Gamma\rho^{(1)}_{DE} \\
%\dot\rho^{(2)}_{DE}+3(1+w)H\rho^{(2)}_{DE}= \Gamma\rho^{(1)}_{DE} \\
%\frac{3H^2}{8\pi G} = \rho^{(1)}_{DE}+\rho^{(2)}_{DE} + 
%\frac{3H_0^2}{8\pi G}(\Omega_{0b} + \Omega_{DM})(1+z)^3,
%\label{eq:model-II}
%\er
%where we assume that $w$ is the EOS of the secondary DE component, $\rho^{(2)}_{DE}$,
%into which the primary component, $\rho^{(1)}_{DE}$, decays, and that $w<-1/3$.
%These eqn's have a free parameter, $w$, and reduce to those of Model II and III when 
%$w = 0, 1/3$ respectively.
%}

\subsection{Model II: Dark Energy decays into Dark Matter} 

The basic equations for a model in which DE decays into dark matter are
\br
\dot\rho_{DE}&=&-\Gamma\rho_{DE} \\
\dot\rho_{DM}+3H\rho_{DM}&=& \Gamma\rho_{DE} \\
\frac{3H^2}{8\pi G}&=& \rho_{DE}+\rho_{DM} + 
\frac{3H_0^2}{8\pi G}\Omega_{\rm 0b}(1+z)^3.
\label{eq:model-II}
\er
where we have assumed that DE, prior to its decay, had the form of a cosmological constant
with $w=-1$. This model formally belongs to the class of interacting DM-DE models
considered in many papers, see~\citep{Amendola:1999er,Guo:2007zk,Boehmer:2008av,Valiviita:2008iv,He:2008tn,Micheletti:2009pk,He:2010im,Pavan:2011xn,Faraoni:2014vra,Salvatelli:2014zta,Abdalla:2014cla} and the recent review \citep{Wang:2016lxa}. 
However, in almost all of these papers the parameter $\Gamma$ was
assumed to be proportional to $H$ or some other time dependent variable 
whereas we assume $\Gamma$ to be a fundamental constant.  
For $\Gamma > 0$ energy flows from the cosmological constant into dark matter (not baryons). It therefore
follows that the current value of $\Omega_{\rm 0m}$, when extrapolated to high redshifts via
$\Omega_{\rm 0m}(1+z)^3$ would be higher than the actual total matter density at
high $z$. This would imply that the expansion rate at high $z$ was {\em lower} that that
in $\Lambda$CDM thereby alleviating some of the tension which exists between 
concordance cosmology and the lower value of $H(z=2.34)$ obtained from quasar-based BAO data
\citep{Delubac:2014aqe}.

%{\bf Varun: We mention $\Omega_{\rm 0m}$ in the para above, but elsewhere in the paper we refer
%only to $\Omega_m$. In case the two are identical then we should replace $\Omega_m$
%by $\Omega_{0m}$ everywhere in the paper. This will be in agreement with our usage of 
%$\Omega_{0b}$.}

\subsection{Model III: Dark Energy decays into Dark Radiation}

For completeness
%, as well with the aim to check if this model can be excluded by the present observational data, 
we also consider a model in which DE decays into ultra-relativistic ``dark'' particles:

\br
\dot\rho_{DE}&=&-\Gamma\rho_{DE} \\
\dot\rho_{DR}+4H\rho_{DR} &=& \Gamma\rho_{DE} \\
\frac{3H^2}{8\pi G} &=& \rho_{DE}+\rho_{DR} + 
\frac{3H_0^2}{8\pi G}\Omega_{\rm 0m}(1+z)^3~.
\label{eq:model-III}
\er

Here the non-relativistic matter component includes both dark matter and baryons, and we have
 neglected the CMB density
$\rho_r$ assuming it to be smaller than that of dark radiation $\rho_{DR}$
(this is plausible for $z \lleq 10$).

\section{Analysis}
\label{sec:analysis}

In order to constrain our three models
 we use combinations of different cosmological data sets, including:

\begin{enumerate}
 
\item
Supernovae Type Ia data from the Union-2.1 compilation containing
 580 Supernovae~\citep{Suzuki:2011hu} within $z\sim0.015-1.4$. We use
 the complete covariance matrix which takes into account systematic effects. 

\item Four BAO datasets: SDSS DR7 ($z=0.35$)~\citep{Percival:2009xn}, BOSS DR9 ($z=0.57$)~\citep{Anderson:2012sa}, 
6DF ($z=0.106$)~\citep{Beutler:2011hx} and SDSS DR11 BAO measurements of $H(z)$ data at $z=2.34$~\citep{Delubac:2014aqe}.

For SDSS DR7 and BOSS DR9 we calculate $D_{\rm V}(z)/r_{s}(z_{\rm drag})$~\footnote{$r_{s}(z_{\rm drag})$ is the comoving 
sound horizon at redshift $z_{\rm drag}$, when baryons decouple from photons.
 $D_{\rm V}=\l[(1+z)^2 D_{\rm A}^2(z)cz/H(z)\r]^{1/3}$ where 
$D_{\rm A}$ is the angular diameter distance.}. 
We do not use the BAO data from WiggleZ because the acoustic parameter $A(z)$ is 
estimated using a specific shape of the power spectrum. The 
WiggleZ data may therefore be biased towards a particular form of the primordial power spectrum and using it could 
bias our overall results.  

\item We include the CMB into our analysis by using the CMB shift parameters $R,l_{a}$ together with the baryon density 
$\Omega_{\rm 0b}h^2$.  The shift parameters are defined as follows: $R=\sqrt{\Omega_{\rm 0m}H_{0}^2}r(z_{*})/c$ and $l_{a}=\pi r(z_{*})/r_s(z_{*})$, 
$r(z_{*})$ being the comoving distance to the photon-decoupling epoch $z_{*}$. 
We use the Planck constrains for these parameters as provided in~\citep{Wang:2013mha}. 
%Note that we only use the shift parameters and the 
%baryon density constraints of ~\citep{Wang:2013mha} here.
\end{enumerate}

For model I we solve the differential equation for $h(z)$ appearing from Eq.~\eqref{eq:rad_de_hz}, get $w(z)$ and use it for further 
evaluation. Since by definition $h(z=0)=1$ we can solve for $h(z)$ just by providing $\Gamma/H_0$ and $\Omega_{\rm 0m}$ for a flat Universe. 
For model II and model III we solve equations~\eqref{eq:model-II} and~\eqref{eq:model-III} to obtain $h(z)$ and the density parameters. Note that in these 
cases we keep the equation of state of the components of the Universe to be same as in the standard case. For model III we assume the dark energy
is coupled to a radiation component and hence we use $w_{\rm DR}(z)=1/3$ for dark radiation.

We work with three combinations of datasets: (i) We use the Union-2.1 compilation jointly with 
 BAO data. (ii) Next we add $H(2.34)$ to (i). (iii) Finally we include the values for the
CMB shift parameters to SNIa, BAO and $H(2.34)$ data. Note that for (i) and (ii)
 we use BBN constraints~\citep{Agashe:2014kda} of $\Omega_{\rm 0b}h^2$ as a prior.
%{\bf Varun: Since we have referred to $\Omega_{\rm 0b}$ earlier, presumably 
%$\Omega_{\rm b}h^2$ should be changed to $\Omega_{\rm 0b}h^2$.}

We use {\tt CosmoMC}~\citep{Lewis:2002ah} to obtain a complete Markov Chain Monte Carlo (MCMC) estimation. Note that for model II and model III, there are certain 
non-physical areas in the parameter spaces \footnote{We have not taken into account the most recent data \citep{Alam:2016hwk} that was released when this paper was being prepared for publication. We expect our results and conclusions will not change considerably by incorporating these data.}.

%Just before finalising this work we faced some theoretical ambiguities that holds us back to recheck all the 
%analysis. In this period there have been some new cosmological data released. We expect our results and conclusion could not be changed considerably by incorporating
%these most recent data.}. 

%For example, for $\Gamma/H_0>0$, as we go back in time, dark radiation decays to dark energy and hence for a very small
%$\Omega_{\rm DR}$ the dark radiation component decays to zero and the evolution breaks down. For such cases, we reject those parameter spaces with providing negligible likelihood.  

%%%%%%%%%%%%%%%%%%%%%%%%%%%%%%%%%%%%%%%%%%%%%%%%%%%%%%%%%%%%%%%%%%%%%%%%%%%%%%%

%%%%%%%%%%%%%%%%%%%%%%%%%%%%%%%%%%%%%%%%%%%%%%%%%%%%%%%%%%%%%%%%%%%%%%%%%%%%%%%
\section{Results}\label{sec:results}

\begin{figure*}
\begin{center} 
\resizebox{220pt}{160pt}{\includegraphics{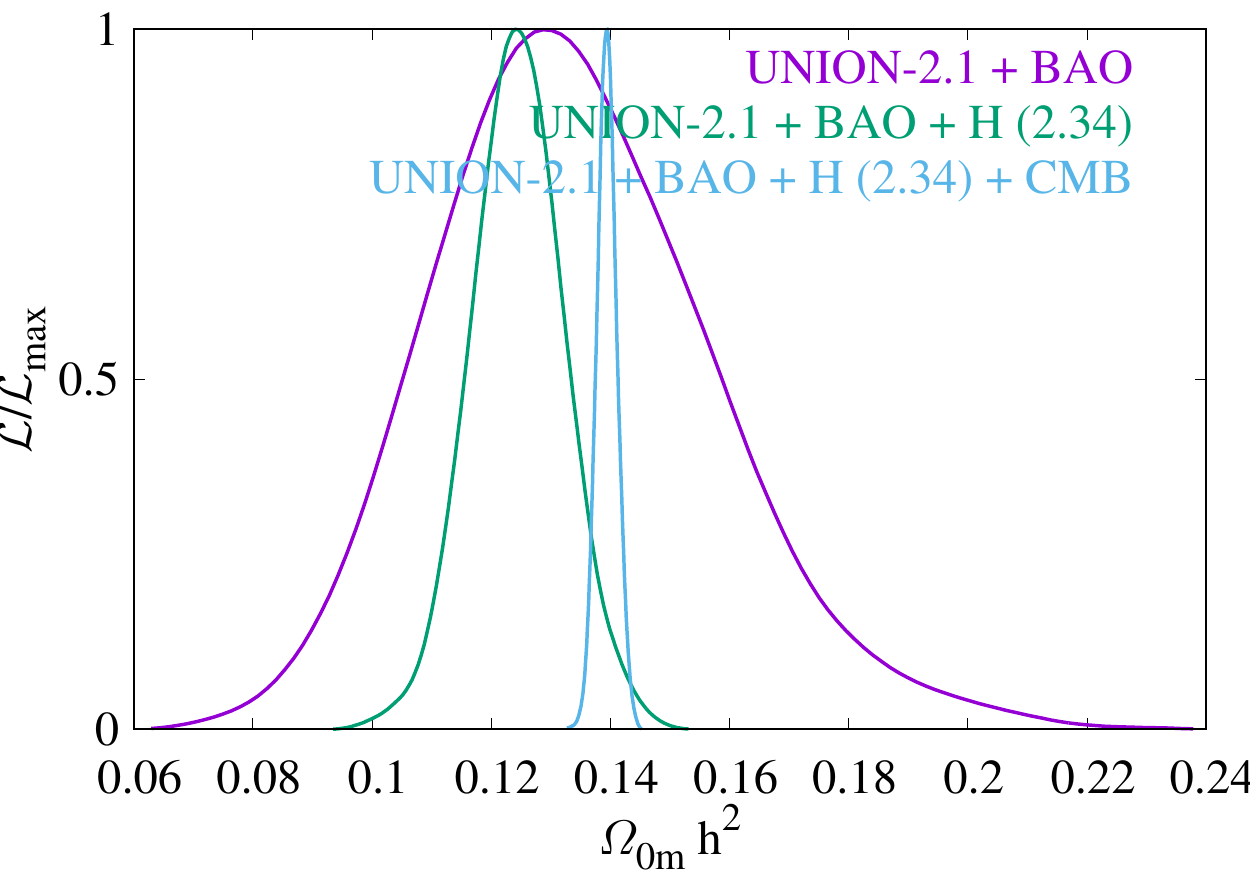}}\hskip -5pt 
\resizebox{220pt}{175pt}{\includegraphics{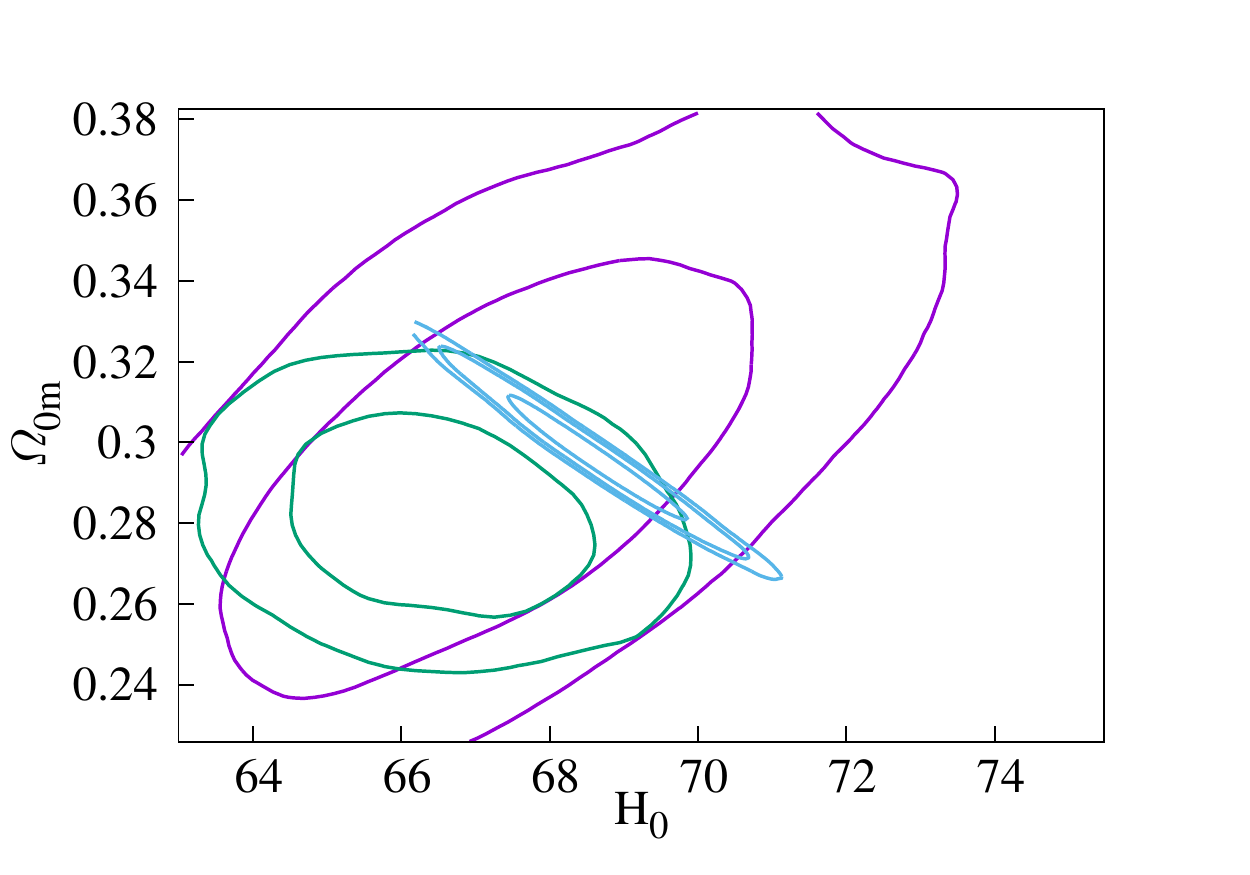}} 
\end{center}
\caption{\label{fig:cntrs-lcdm} Observational constraints on standard $\Lambda$CDM. [Left] One dimensional marginalized likelihoods of $\Omega_{\rm 0m}h^2$. [Right] $\lbrace\Omega_{\rm 0m},H_0\rbrace$ contours for different datasets. The color-code convention for the right panel is identical to the left.}
\end{figure*}

In order to gain a better understanding as to how our models might provide good fits to the observational data,
we begin by showing results for concordance cosmology ($\Lambda$CDM). In figure~\ref{fig:cntrs-lcdm} (left panel) we show
the 1D marginalized likelihood of $\Omega_{\rm 0m}h^2$ obtained using different data combinations. 
In the right panel the 2D marginalized contours of $\Omega_{\rm 0m}$ vs $H_0$ are shown. 
Constraints on different cosmological parameters are also given in table~\ref{tab:chi2lcdm}. We clearly see that 
the $H(2.34)$ data point pushes the best-fit $\Lambda$CDM model towards a lower matter density as well as 
a lower Hubble parameter. Including the CMB (in combination with other data sets) pushes the best-fit
 back to a higher matter density and a higher Hubble parameter. This tension had earlier been reported in~\citep{Sahni:2014ooa,Delubac:2014aqe}. 
One of the aims of the present analysis is to see whether one can alleviate this tension using our new DE models.\\

\begin{figure*}
\begin{center} 
\resizebox{220pt}{150pt}{\includegraphics{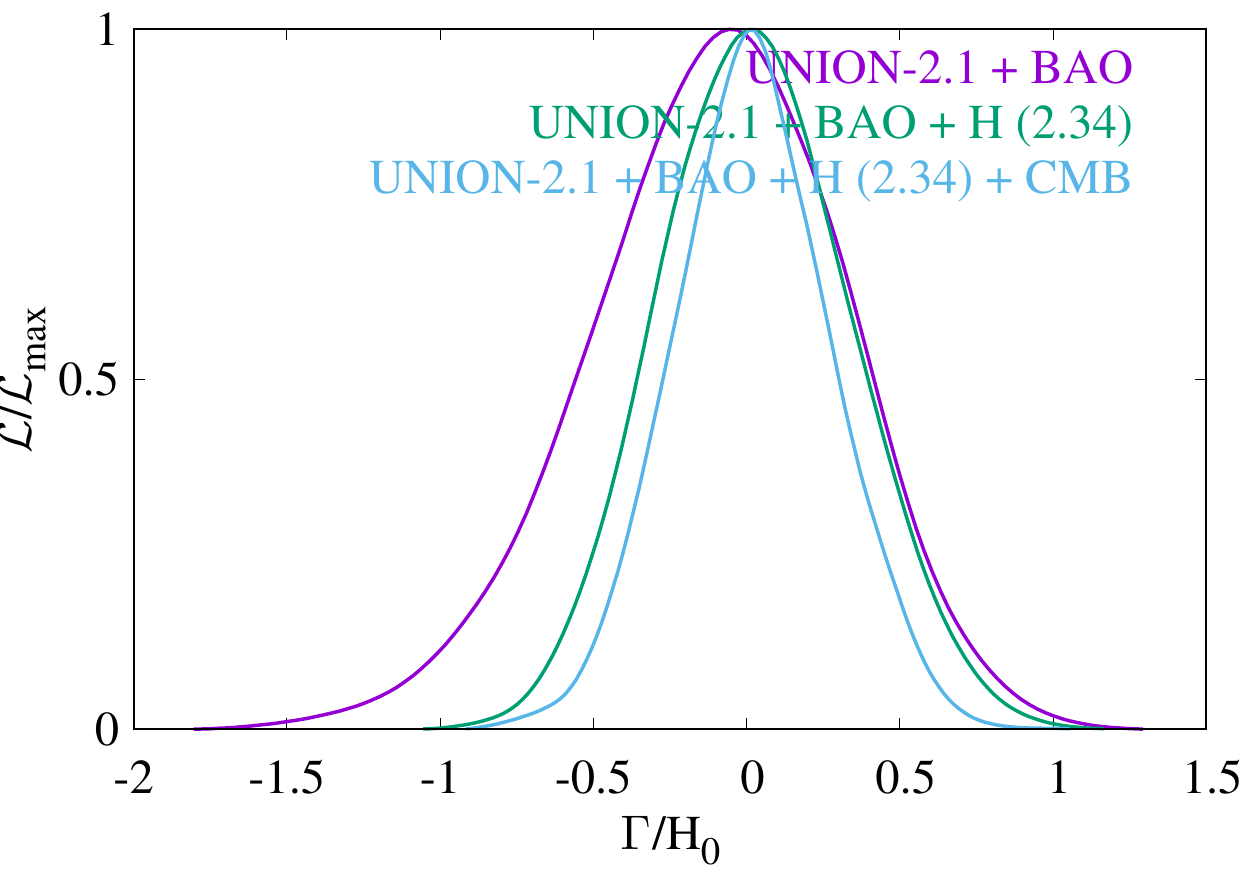}}\hskip -5 pt 
\resizebox{220pt}{150pt}{\includegraphics{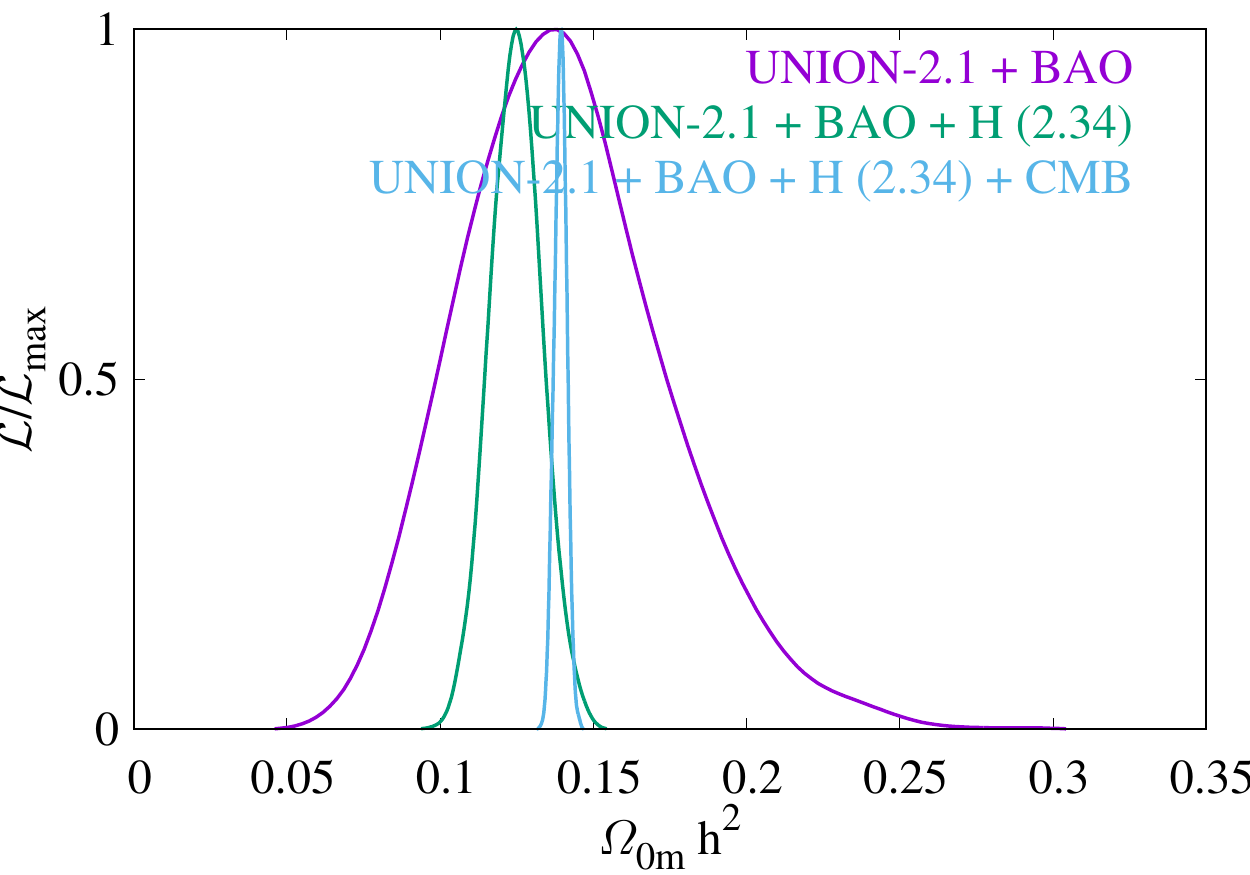}} 

\resizebox{220pt}{180pt}{\includegraphics{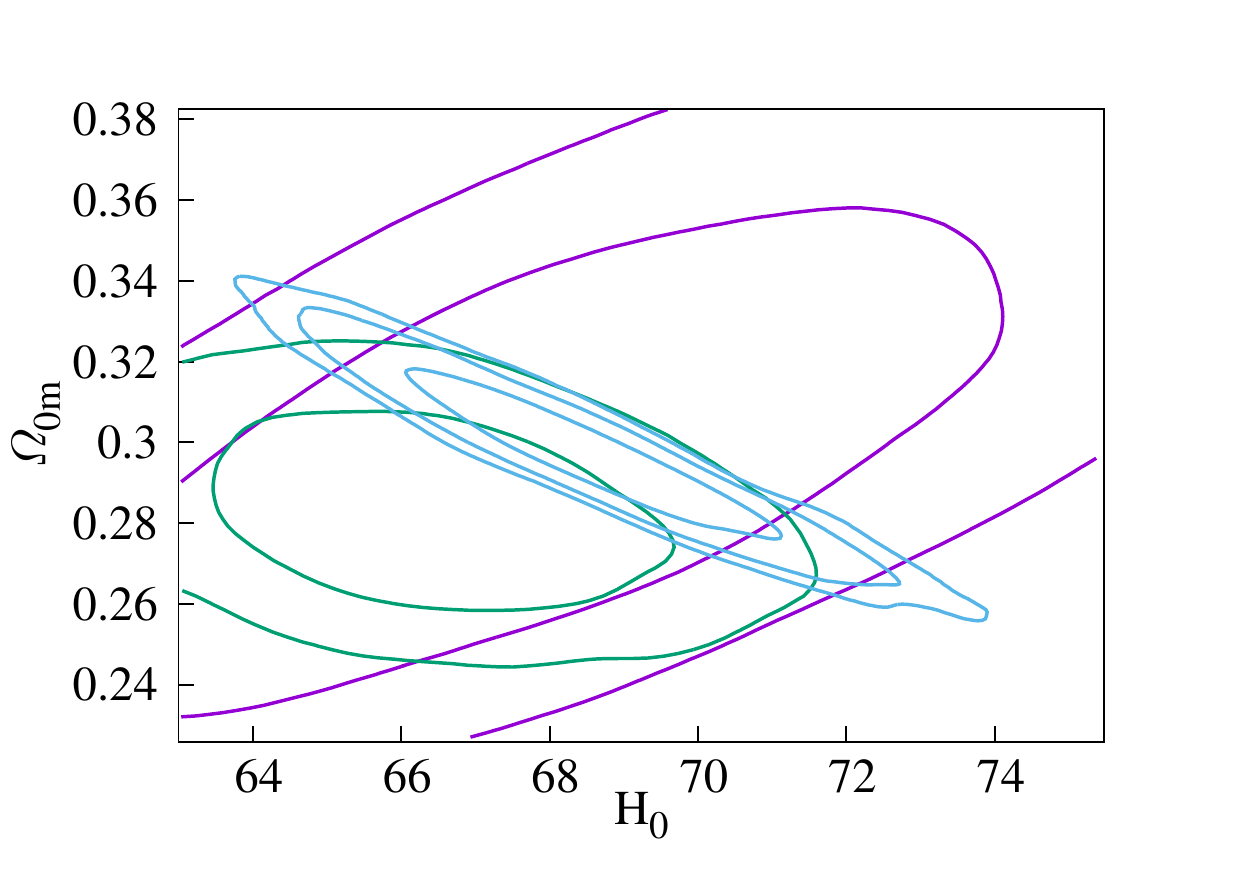}} \hskip -5 pt
\resizebox{220pt}{180pt}{\includegraphics{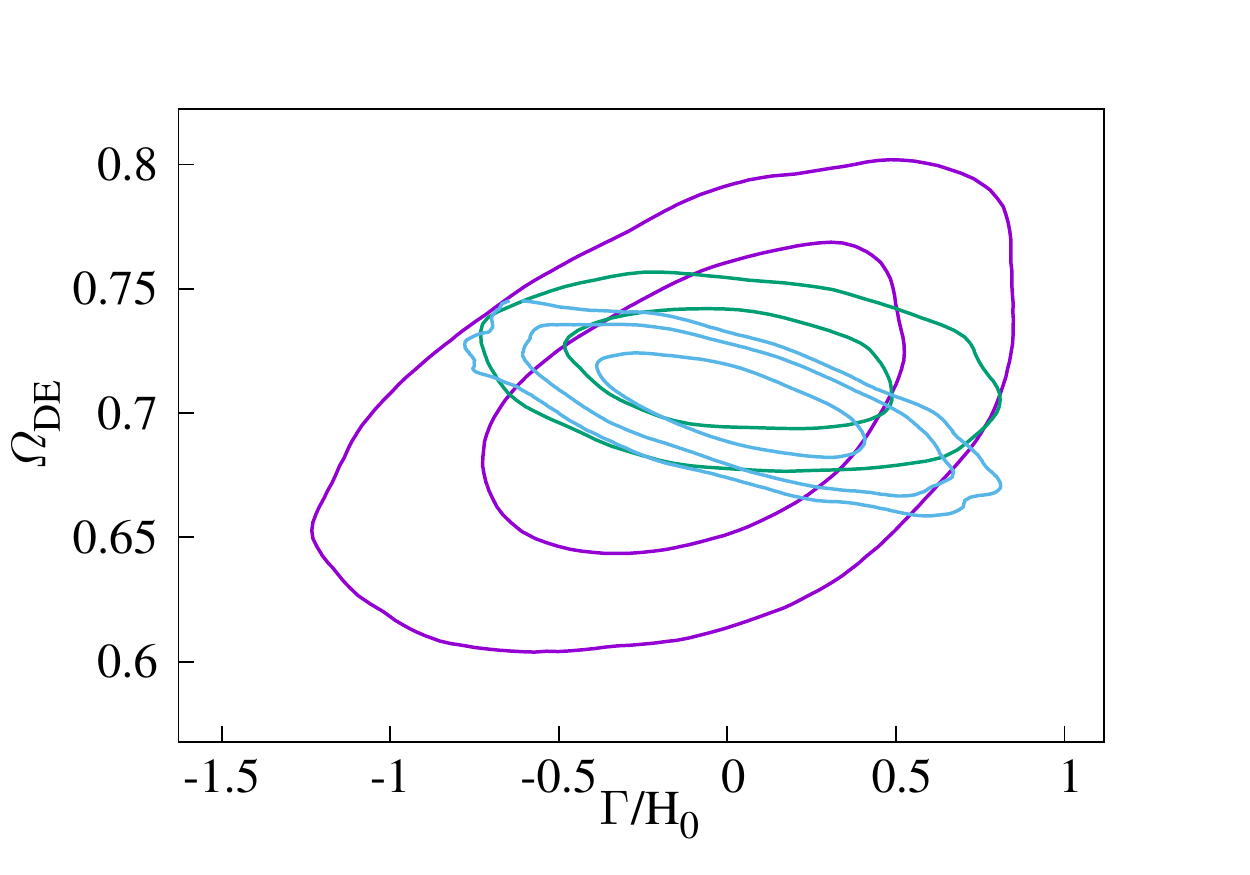}} 

\end{center}
\caption{\label{fig:cntrs-model-I} Results for Model I. [Top left] One dimensional marginalized likelihoods of the decay parameter in Eq.~\ref{eq:model-I}. [Top right] 1D likelihoods of $\Omega_{\rm 0m}h^2$. [Bottom left] 2D contours of $\lbrace\Omega_{\rm 0m},H_0\rbrace$. [Bottom right] 2D contours of $\lbrace\Omega_{\rm DE},\Gamma/H_0\rbrace$. One can see that the tension between CMB data and the $H(2.34)$ data point is somewhat reduced in this model.}
\end{figure*}

In figure~\ref{fig:cntrs-model-I} we show the results for Model I (exponentially decaying dark energy). In the top-left panel we plot the marginalized 1D 
likelihood of the decay parameter $\Gamma/H_0$ obtained from different combinations of datasets. 
In the top-right panel we show the 1D marginalized likelihood for $\Omega_{\rm 0m}h^2$ while
 in the bottom-left panel we show the marginalized contour of $\Omega_{\rm 0m}$ vs $H_0$ 
(these two plots can be compared with the corresponding results for $\Lambda$CDM shown in figure \ref{fig:cntrs-lcdm}). 
In the bottom-right panel we show the marginalized contour of $\lbrace\Omega_{\rm DE},\Gamma/H_0\rbrace$. 
Our results indicate that the presence of the additional degree of freedom, $\Gamma$,
increases the area of the confidence contours and reduces the tension between CMB data and the $H(2.34)$ data point present in $\Lambda$CDM. 
%Also, we can see that such a model, where dark energy as a substance with an internal clock decay to itself can be a viable model even though its half-life must be more than the age of the universe. 
However, the fact that the standard $\Lambda$CDM model lies close to the centre of the marginalized likelihood contours 
($\Gamma/H_0 = 0 $) suggests that while Model I is viable, it is not strongly preferred over concordance cosmology.

From figure~\ref{fig:cntrs-model-I} we can see that both $\Gamma > 0$ and $\Gamma < 0$ are permitted
by the data. As we mentioned earlier, a negative values of $\Gamma$ imply an increase
in the DE density with time which corresponds to phantom-like behaviour with $w_{\rm eff}< -1$. In other words dark energy at late universe would have more density in comparison to the earlier times (at higher redshifts). $\Gamma > 0$ implies decreasing in the DE density with time that can be effectively correspond to quintessense-like behaviour with $w_{\rm eff} > -1$. We should also note that in both cases of $\Gamma > 0$ and $\Gamma < 0$, the effective equation of state of dark energy cannot be a constant value and it would vary by time.

\begin{figure*}
\begin{center} 
\resizebox{145pt}{120pt}{\includegraphics{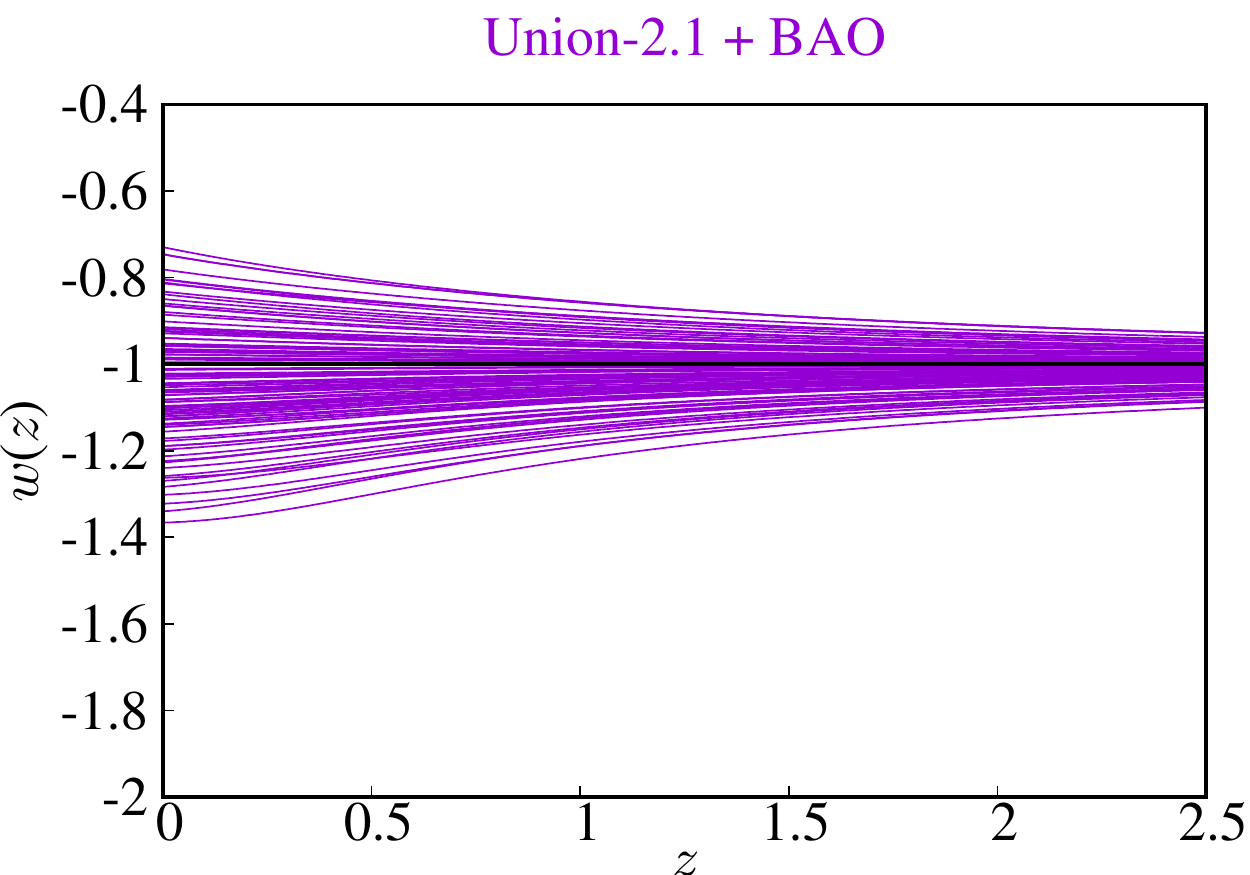}} 
\hskip -5pt \resizebox{145pt}{120pt}{\includegraphics{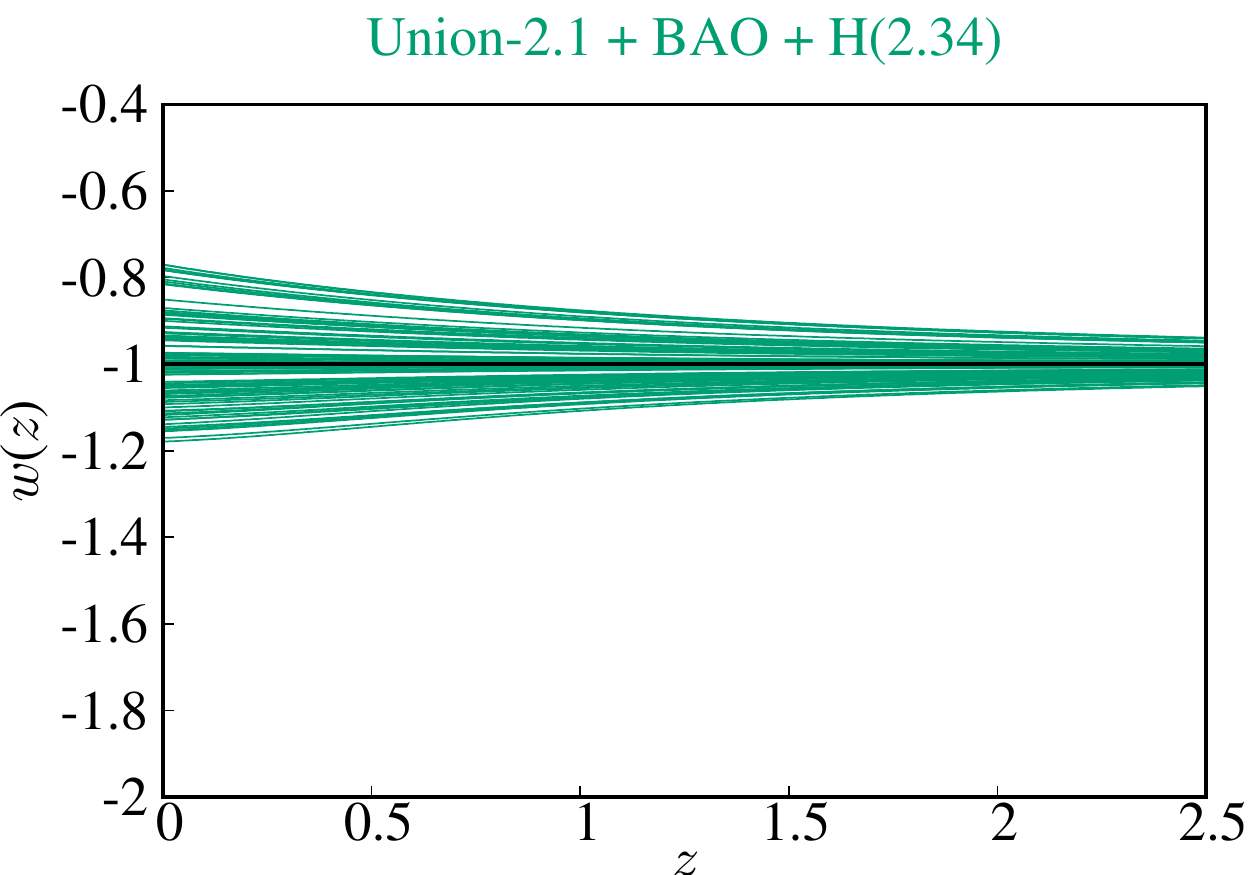}} 
\hskip -5pt \resizebox{145pt}{120pt}{\includegraphics{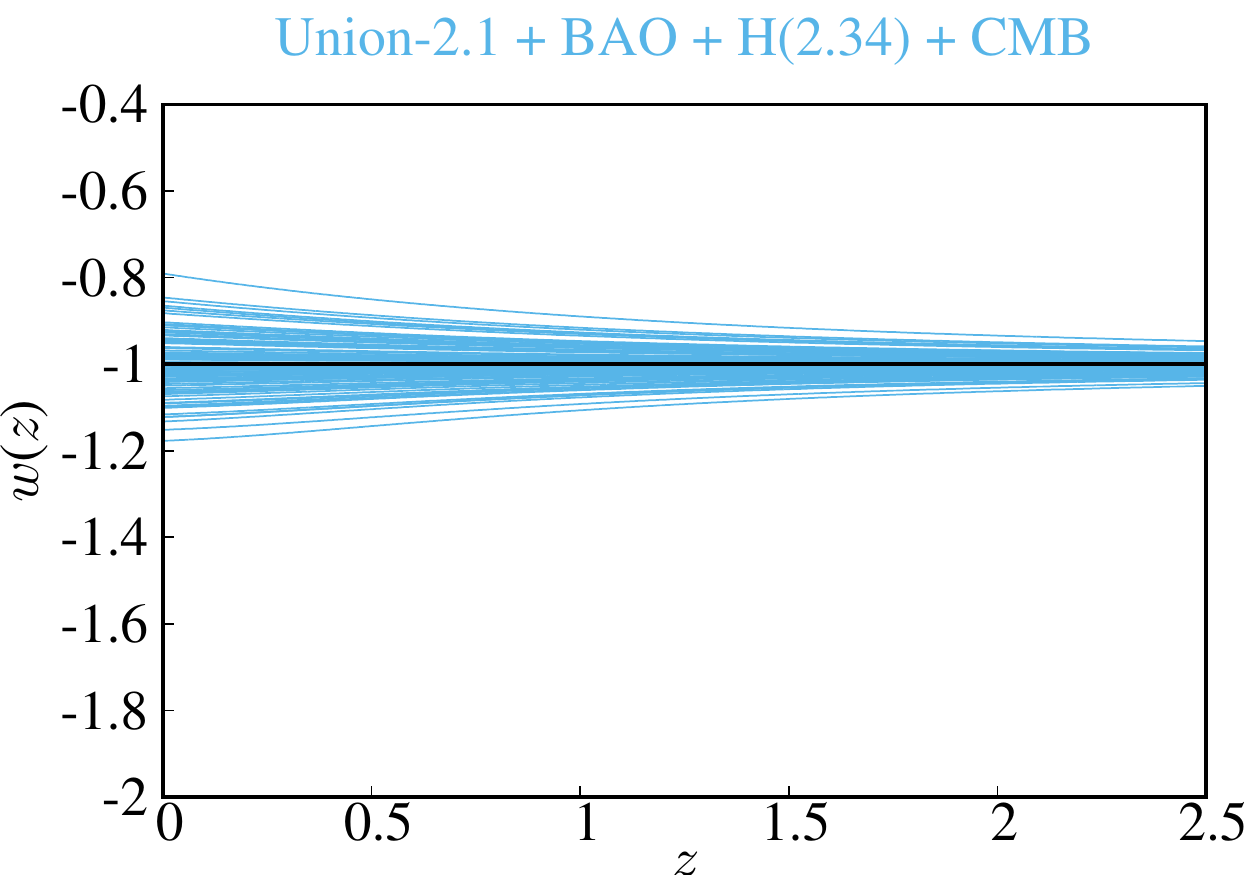}} 

\resizebox{145pt}{120pt}{\includegraphics{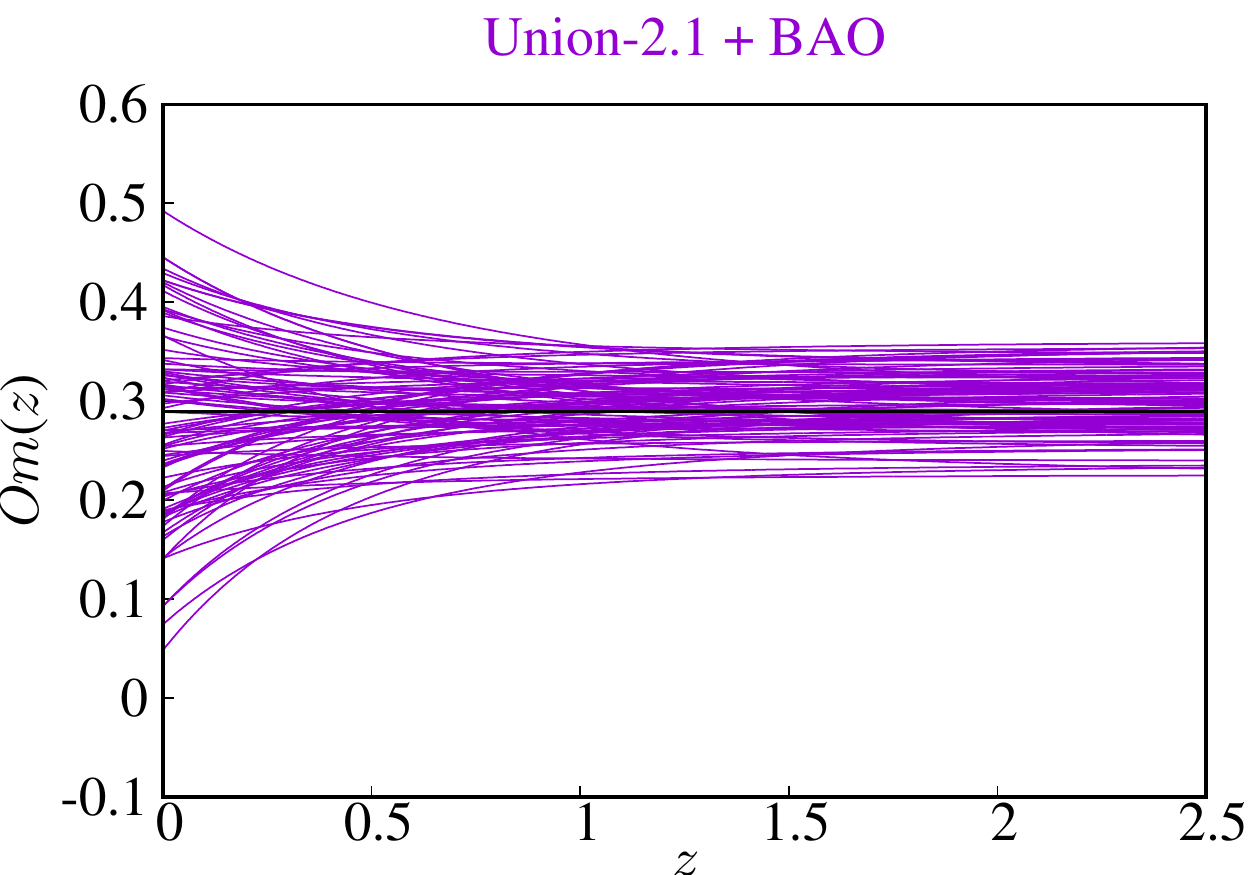}} 
\hskip -5pt \resizebox{145pt}{120pt}{\includegraphics{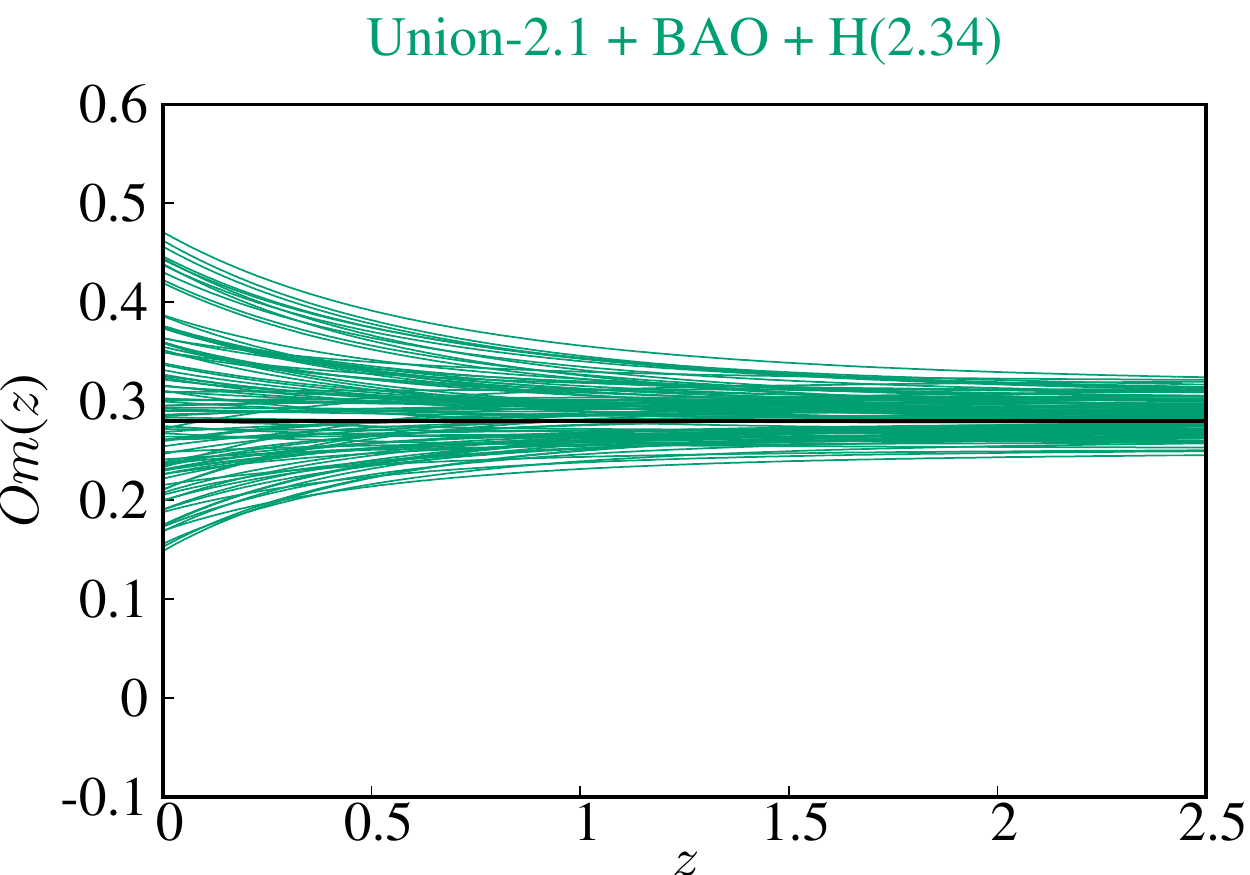}} 
\hskip -5pt \resizebox{145pt}{120pt}{\includegraphics{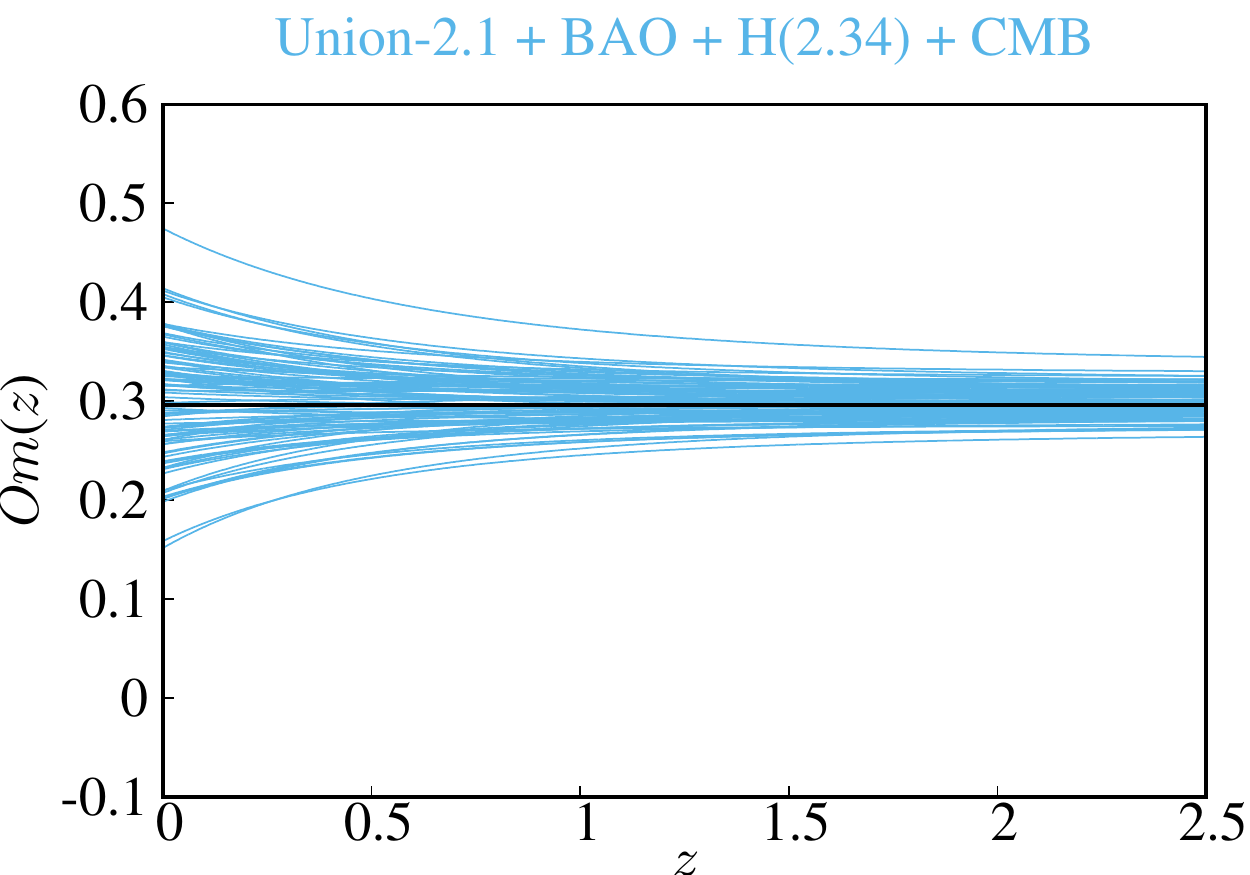}} 

\resizebox{145pt}{120pt}{\includegraphics{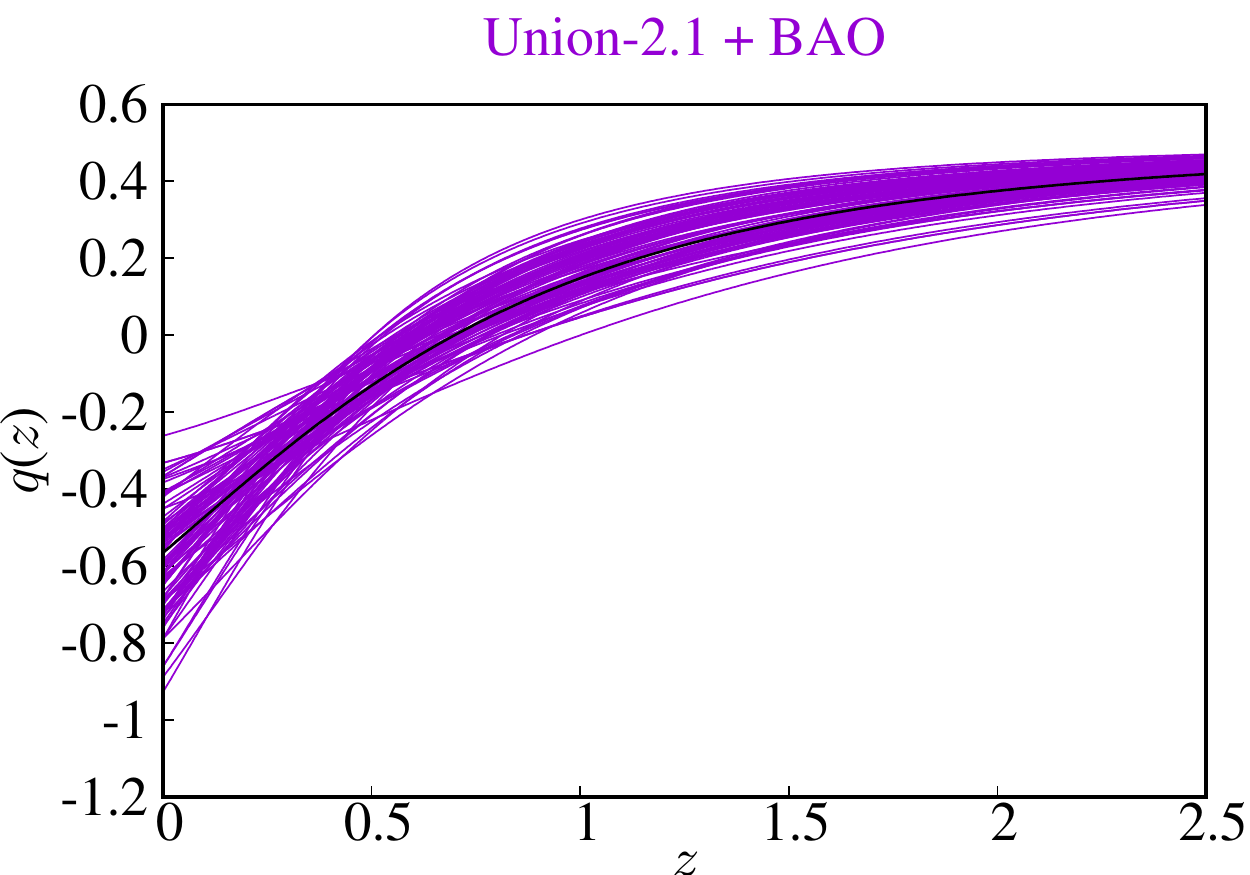}} 
\hskip -5pt \resizebox{145pt}{120pt}{\includegraphics{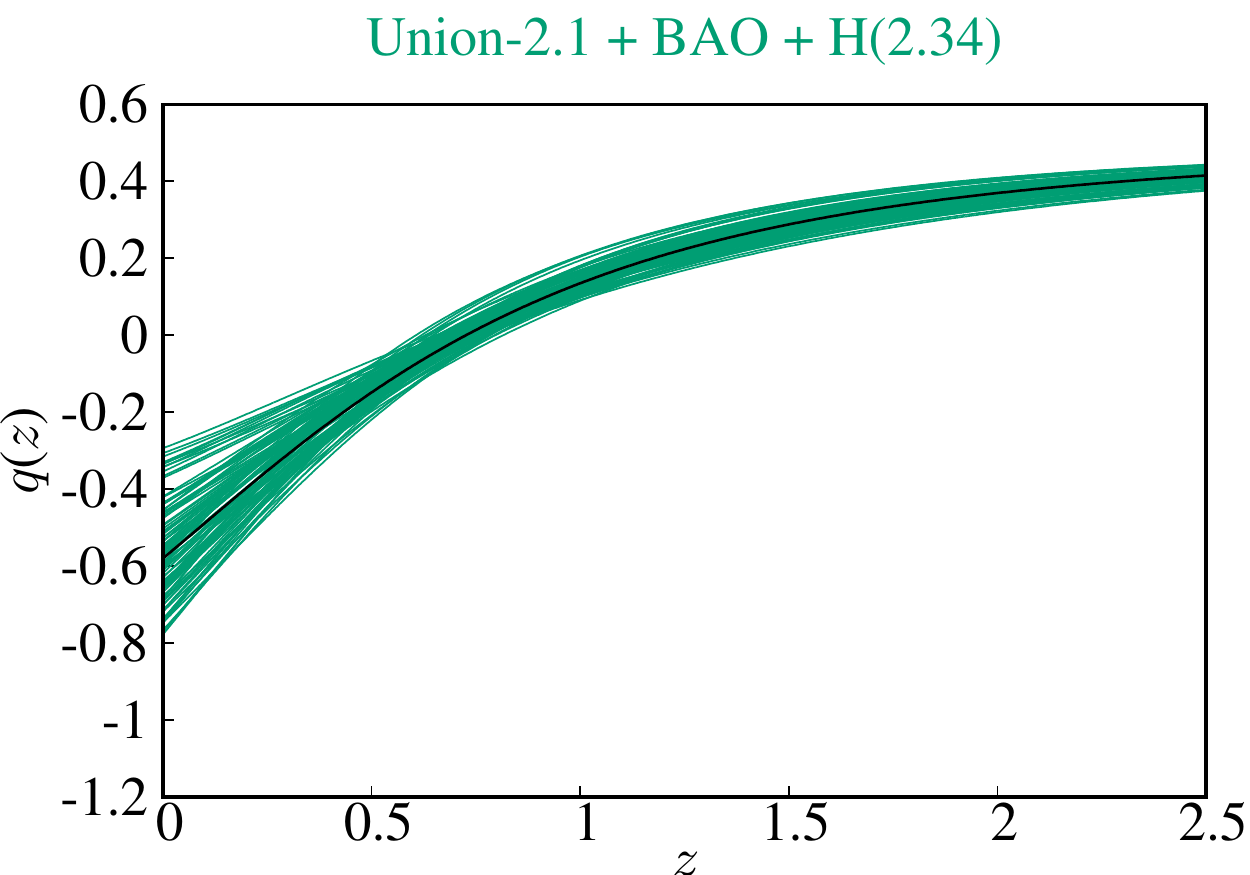}} 
\hskip -5pt \resizebox{145pt}{120pt}{\includegraphics{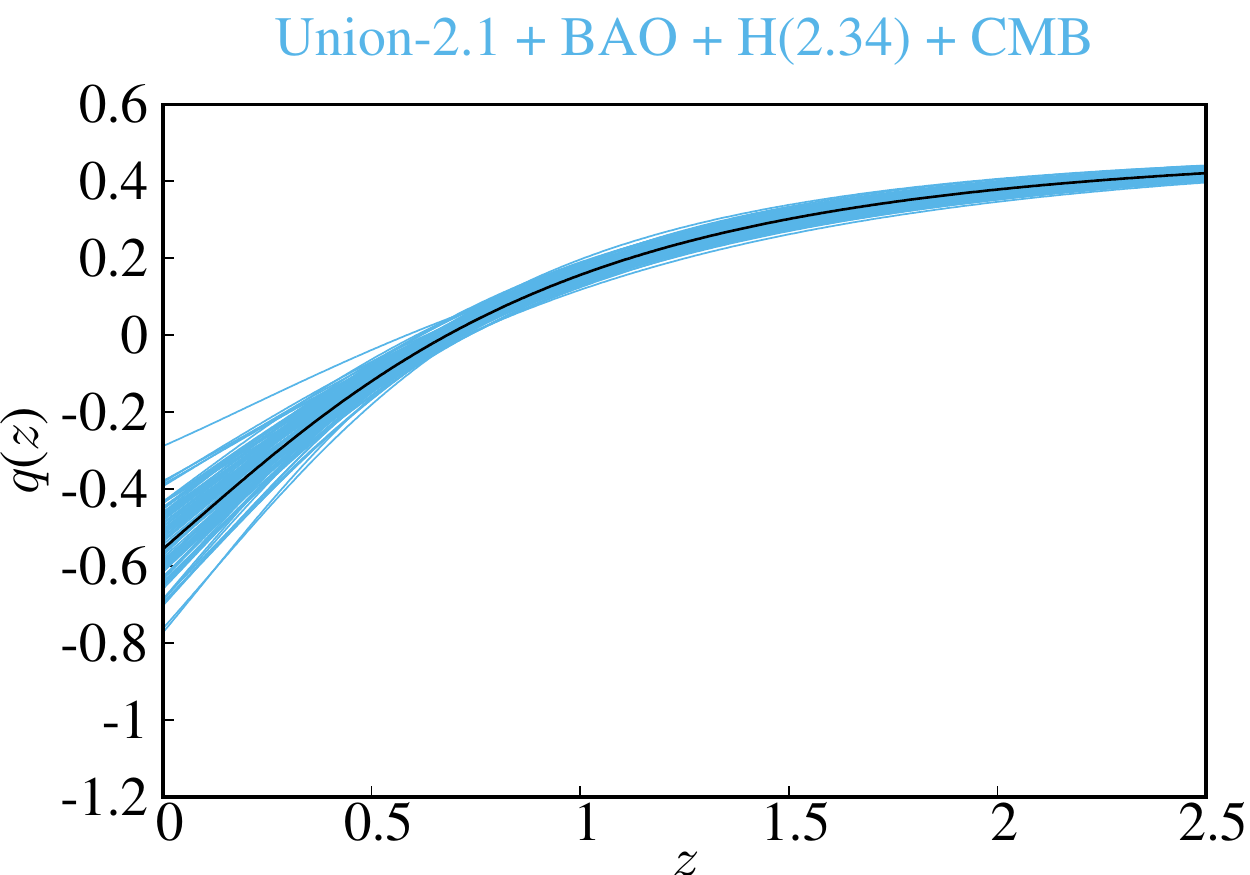}} 

\end{center}
\caption{\label{fig:sample-model-I} The equation of state of dark energy, $w(z),$
the $Om$ diagnostic $Om(z)$, and the deceleration parameter $q(z)$ (top, middle, bottom) are shown as a function of redshift for model I (Eq.~\ref{eq:model-I}). Left to right 
we plot samples within 2$\sigma$ CL's obtained from MCMC chains corresponding to Union-2.1 + BAO,  Union-2.1 + BAO + H(2.34) and  Union-2.1 + BAO + H(2.34) + CMB 
respectively. The black lines correspond to the best fit $\Lambda$CDM for the same combination of datasets.}
%{\bf Varun: The axis labels are very small and should be increased in size. This also applies to the other figures showing $Om$, and $q(z)$.}
\end{figure*}

In figure~\ref{fig:sample-model-I} we show a few important quantities characterizing the expansion of the Universe. 
Top to bottom are shown: (a) the equation of state of dark energy $w(z)$, (b) the  $Om$ diagnostic,
$Om(z)=(h^2(z)-1)/\l[(1+z)^3-1\r]$~\citep{Sahni:2008xx,Shafieloo:2012rs}, (c) the deceleration parameter 
$q(z)=-\dot{H}/H^2-1$.
We plot 100 samples for these three quantities uniformly chosen from within the 2$\sigma$ range of the 
MCMC chains corresponding to
different datasets. It is interesting that, for some parameter combinations, the slowing-down of cosmic
 acceleration appears to be consistent with the data.

\begin{figure*}
\begin{center} 
\resizebox{220pt}{150pt}{\includegraphics{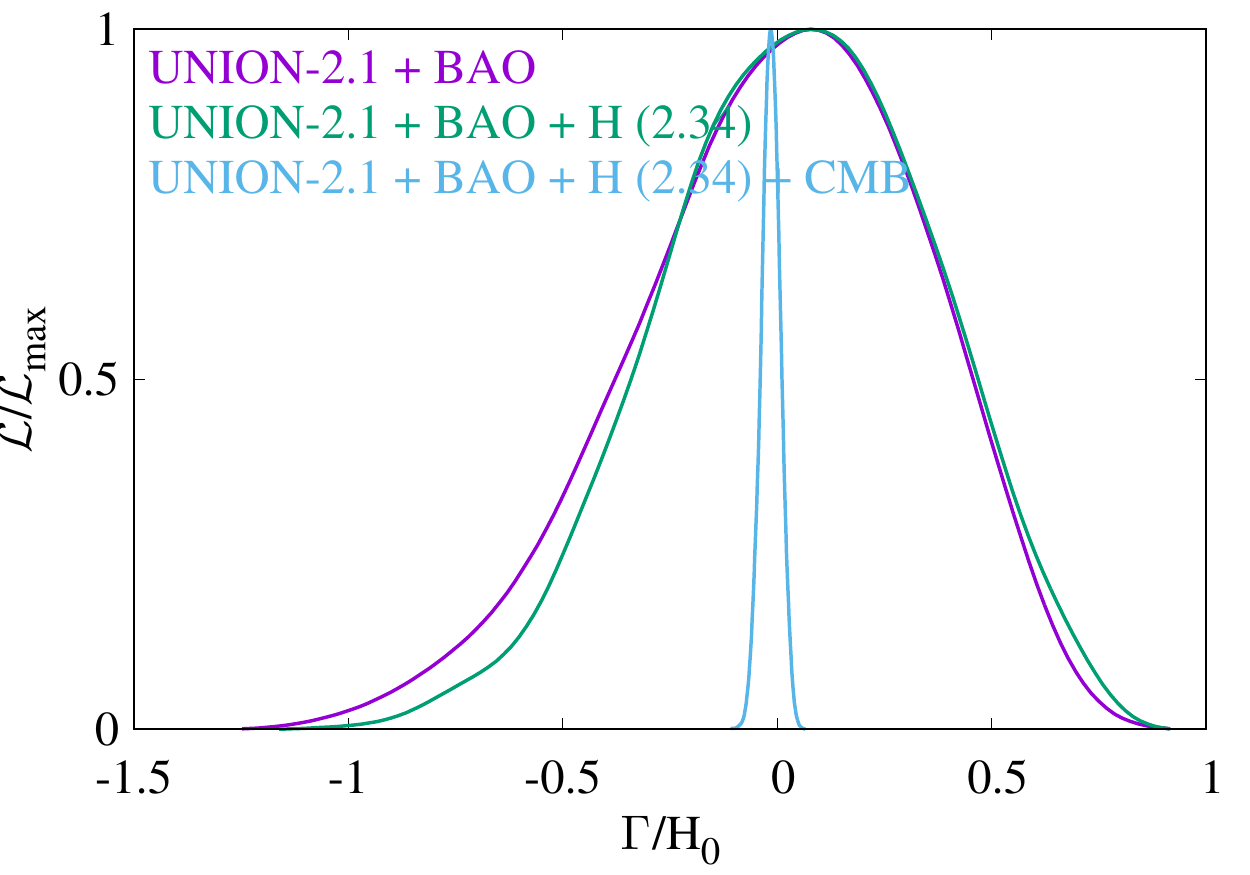}}\hskip -5 pt 
\resizebox{220pt}{150pt}{\includegraphics{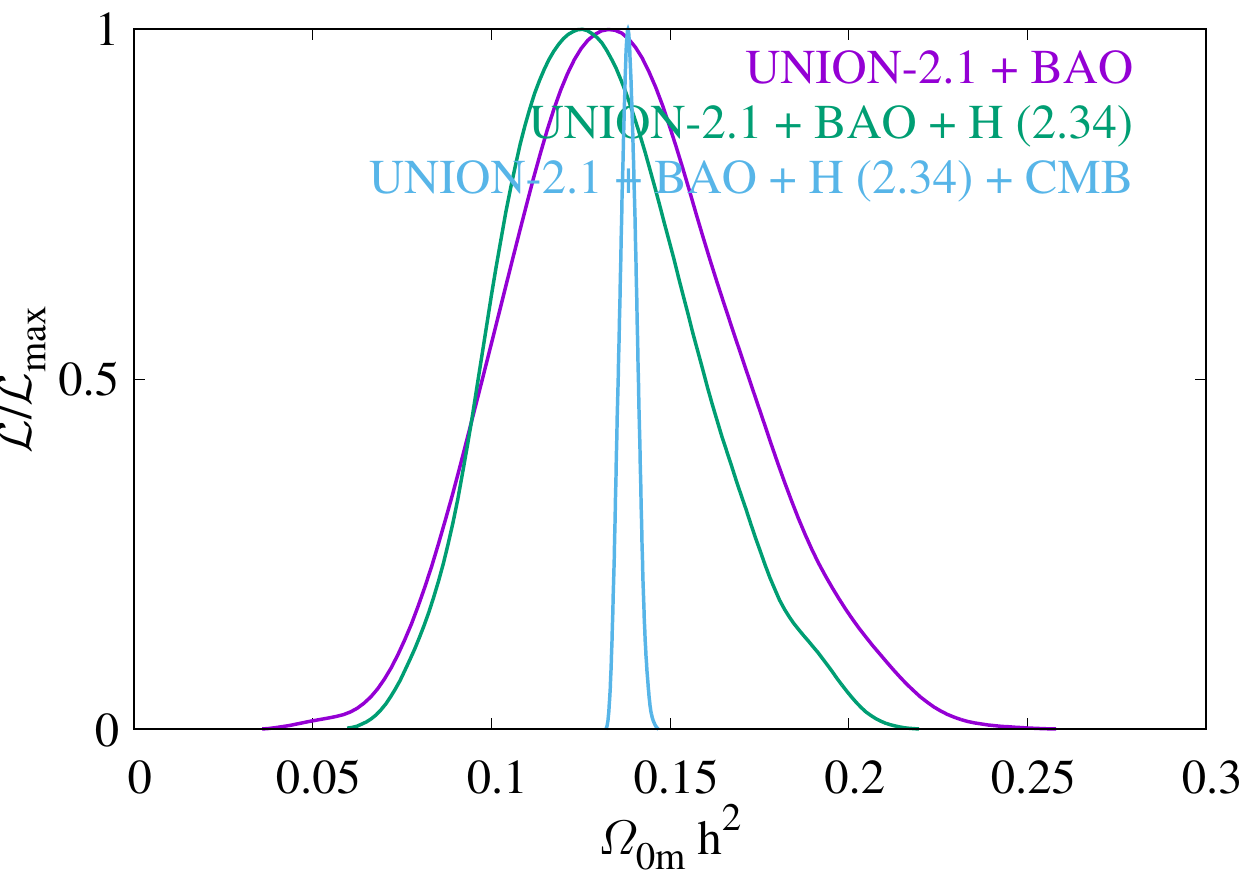}} 

\resizebox{220pt}{180pt}{\includegraphics{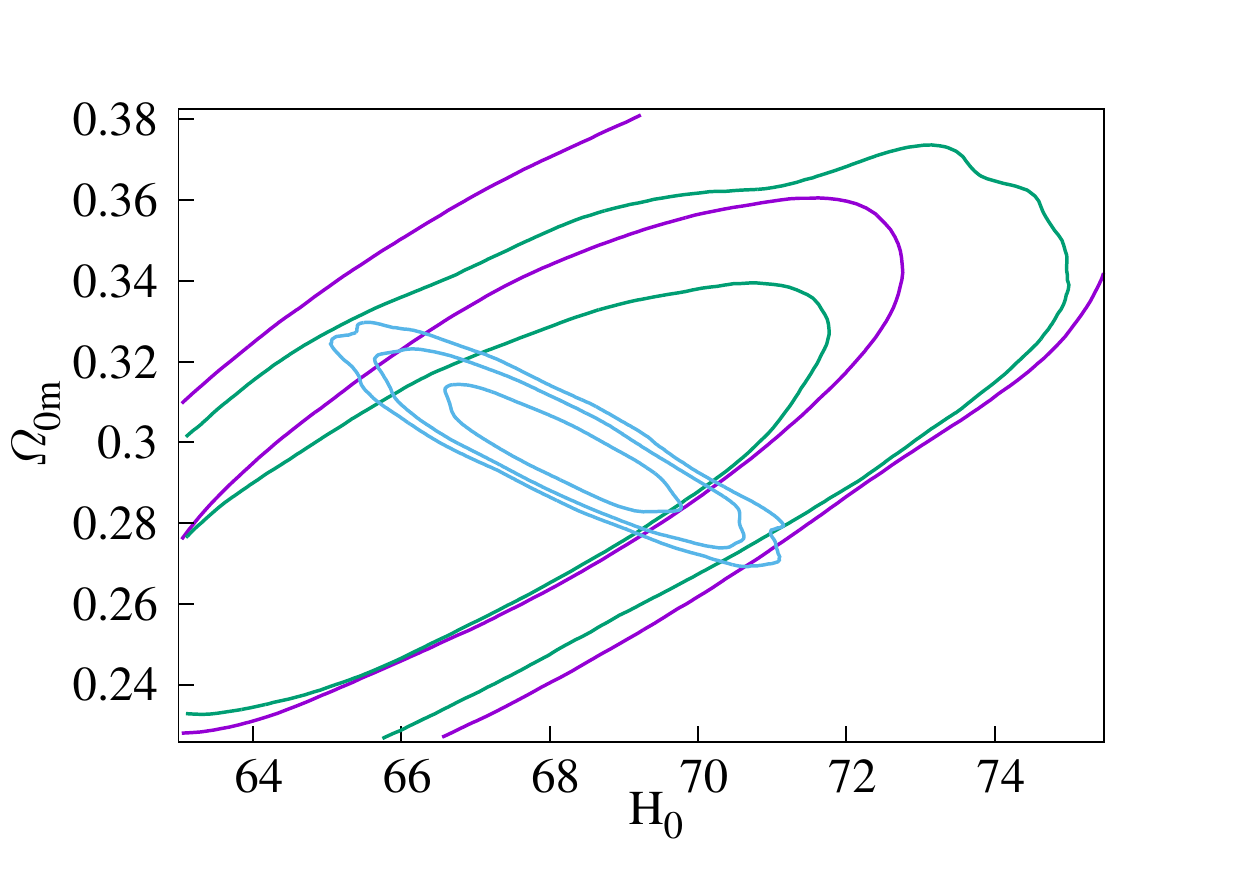}} \hskip -5 pt
\resizebox{220pt}{180pt}{\includegraphics{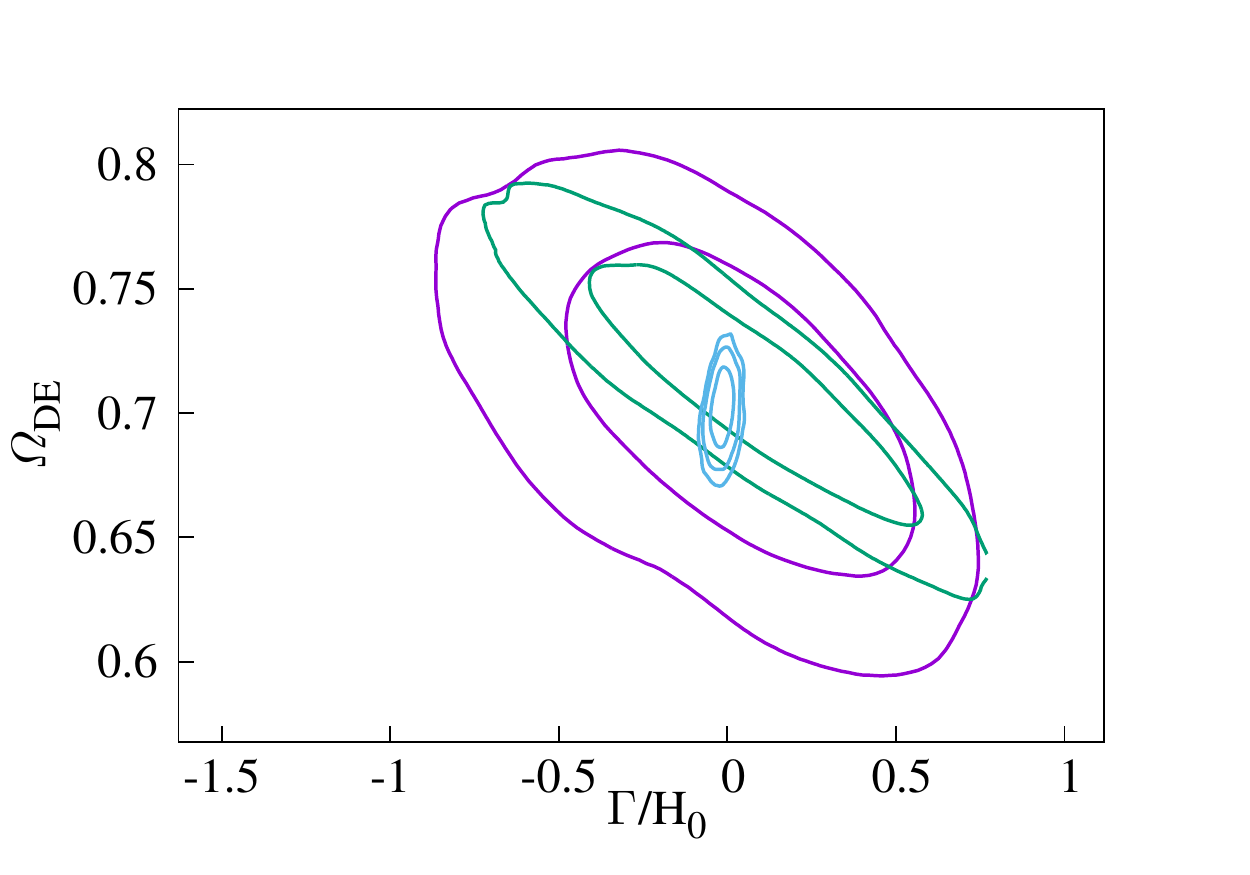}} 

\end{center}
\caption{\label{fig:cntrs-model-II} Results for Model II. [Top left] One dimensional marginalized likelihoods 
of the decay parameter $\Gamma$ are shown for model II (Eq.~\ref{eq:model-II}). 
[Top right] 1D likelihoods of $\Omega_{\rm 0m}h^2$. [Bottom left] 2D contours of $\lbrace\Omega_{\rm 0m},H_0\rbrace$. [Bottom right] 2D contours of $\lbrace\Omega_{\rm DE},\Gamma/H_0\rbrace$. Unlike $\Lambda$CDM, no tension between CMB data and the
$H(2.34)$ data point is indicated for Model II.}
\end{figure*}

\begin{figure*}
\begin{center} 

\resizebox{145pt}{120pt}{\includegraphics{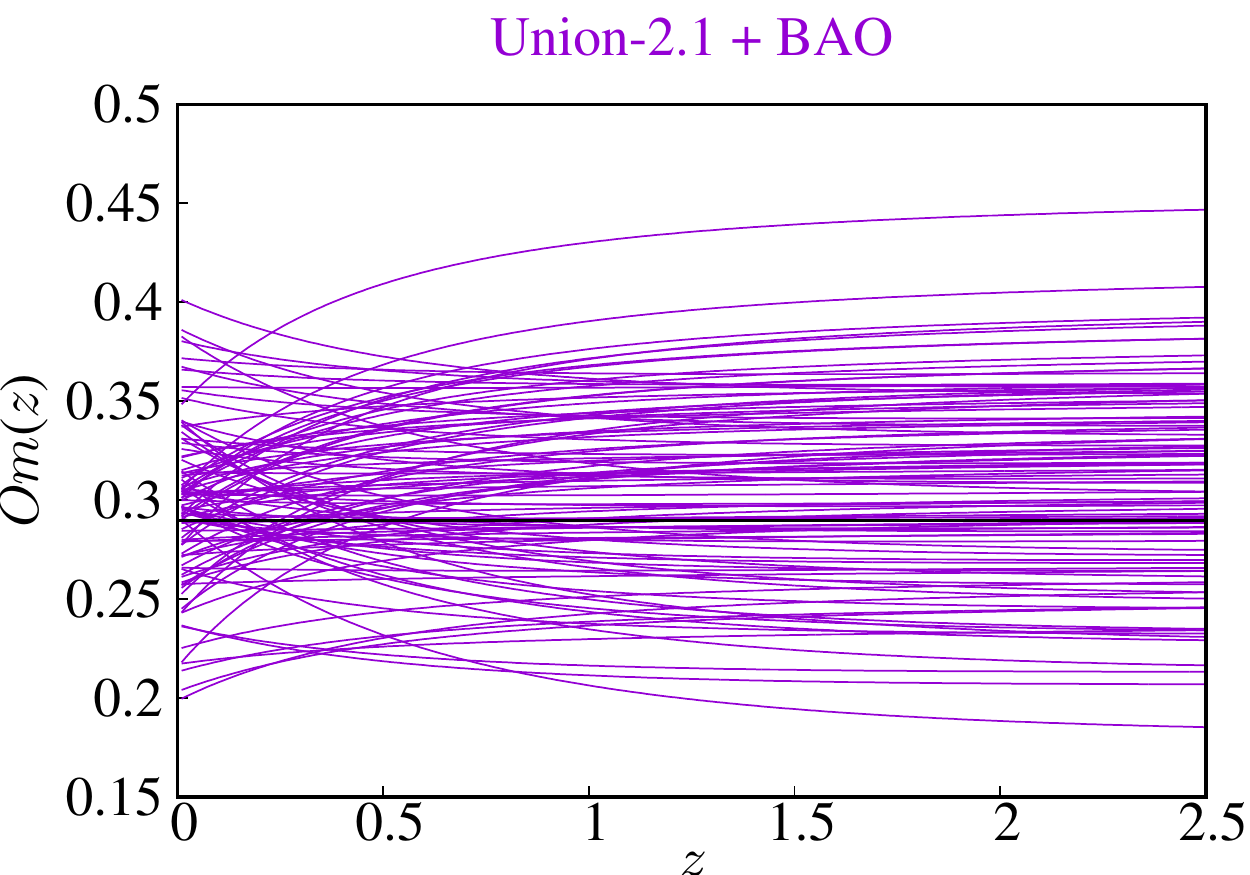}} 
\hskip -5pt \resizebox{145pt}{120pt}{\includegraphics{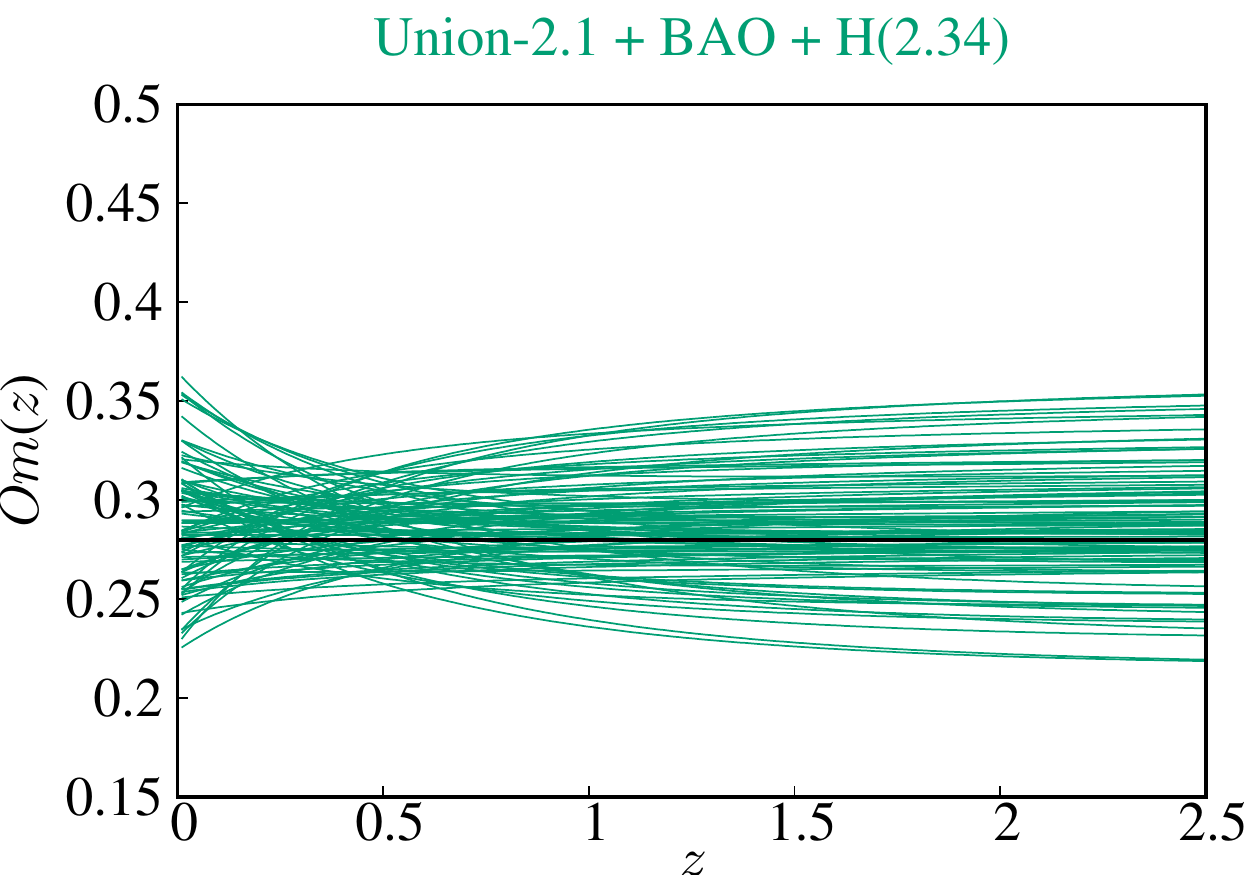}} 
\hskip -5pt \resizebox{145pt}{120pt}{\includegraphics{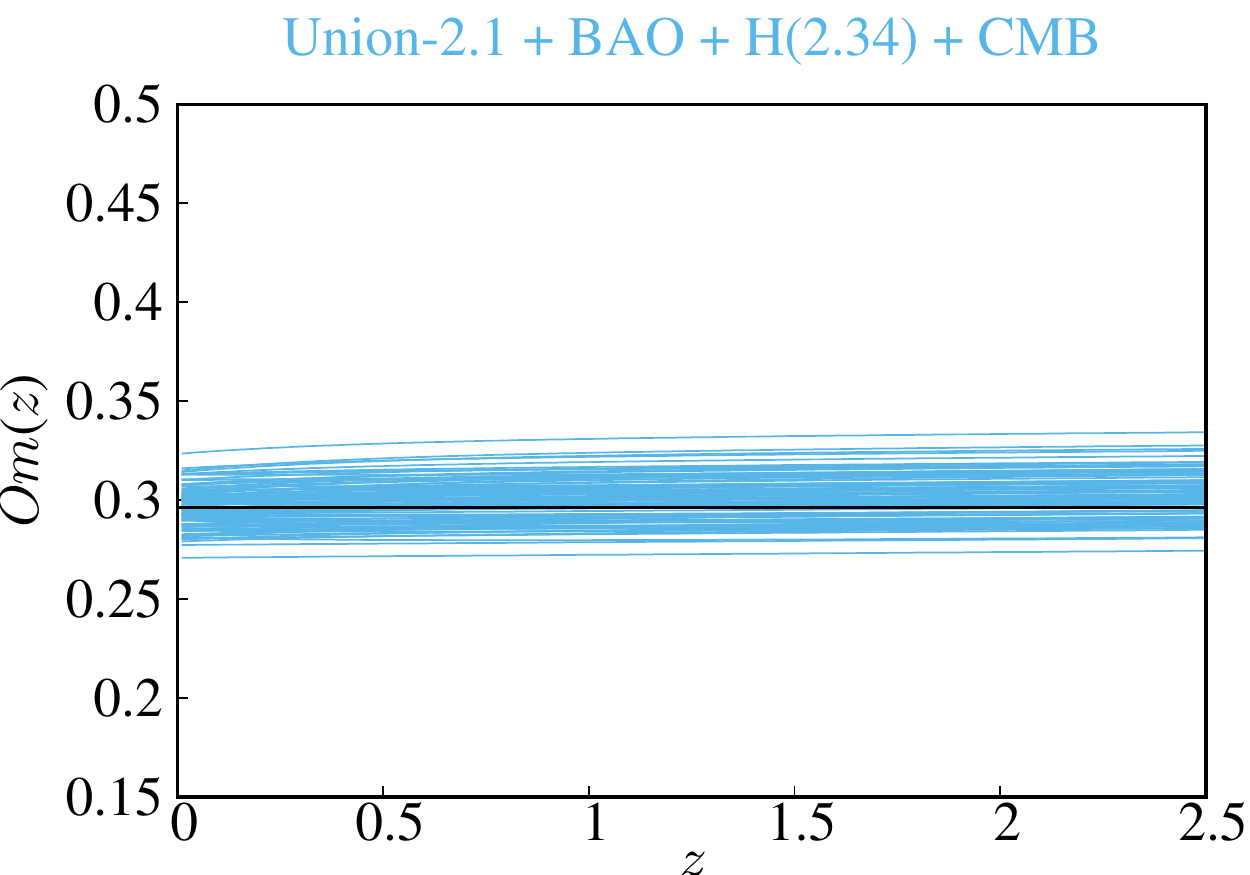}} 

\resizebox{145pt}{120pt}{\includegraphics{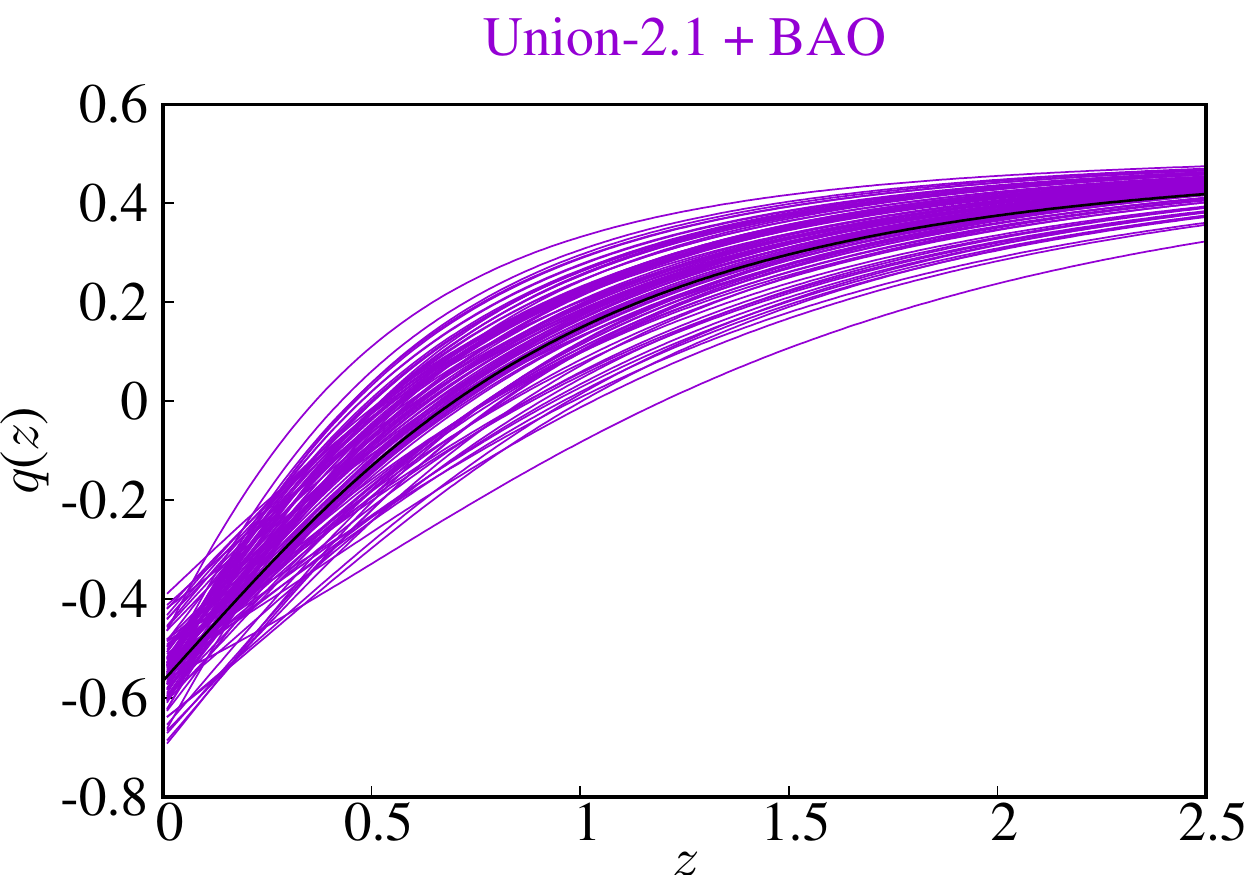}} 
\hskip -5pt \resizebox{145pt}{120pt}{\includegraphics{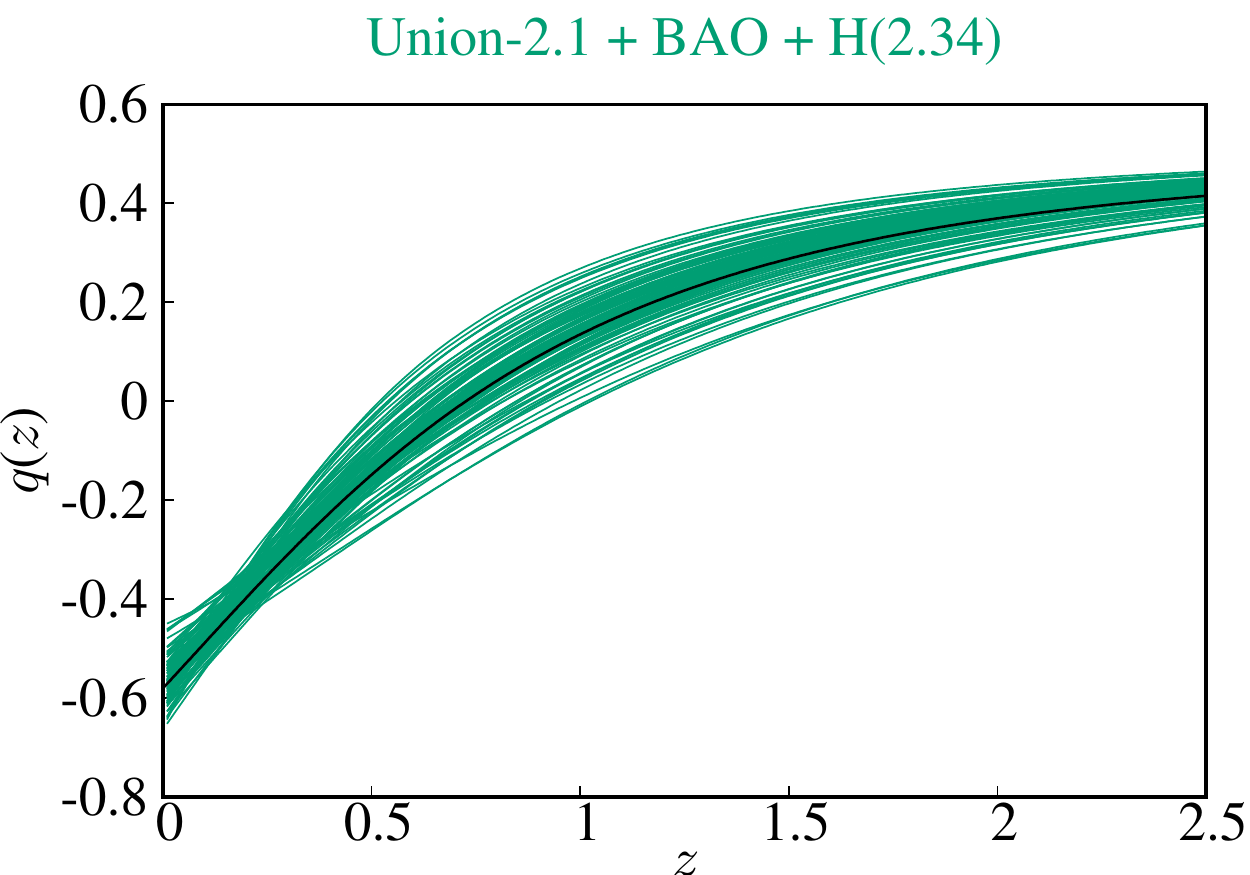}} 
\hskip -5pt \resizebox{145pt}{120pt}{\includegraphics{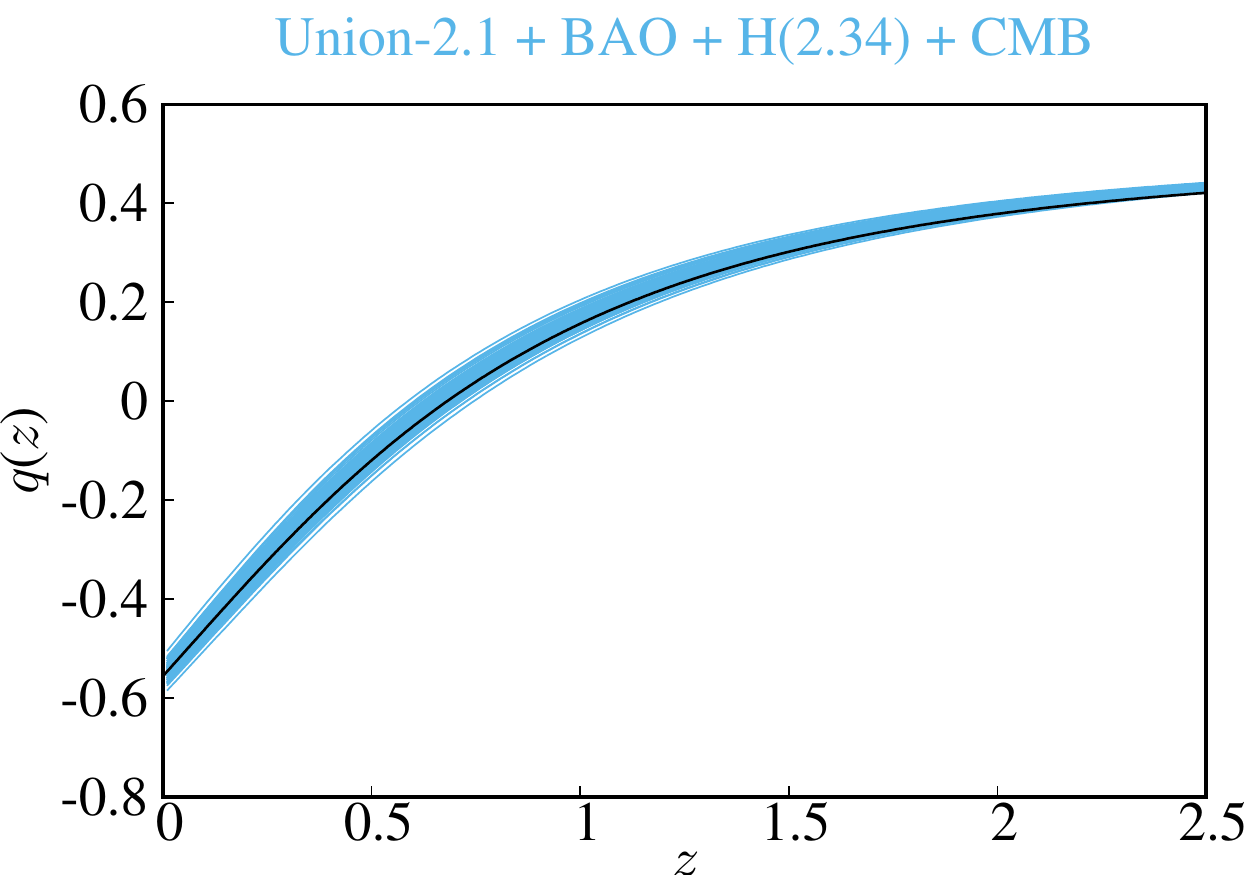}} 
\end{center}
\caption{\label{fig:sample-model-II} Samples of $Om(z)$ and $q(z)$ for model II (Eq.~\ref{eq:model-II}). The datasets are 
indicated at the top of each plot. The black lines correspond to the best fit $\Lambda$CDM for the same combination
of datasets.}
\end{figure*}

In figures~\ref{fig:cntrs-model-II} and \ref{fig:sample-model-II}  we show our 
results for Model II (DE decaying into dark matter). Our results show that in this case
 there is no significant tension between the $H(2.34)$ data point and CMB data.
In fact Model II allows for a wide range of cosmological parameters to be consistent with the data. 
One should note that,
unlike $\Lambda$CDM, there is a large overlap of confidence contours at the $1\sigma$ level
when plotted with and without CMB data. However, as in the case of Model I, $\Lambda$CDM lies close 
to the centre of the confidence contours, which is an indication of how well concordance cosmology is performing. From figure~\ref{fig:cntrs-model-II} we find that both $\Gamma > 0$ and $\Gamma < 0$ are permitted
by the data. $\Gamma > 0$  implies the transfer of energy from dark energy into dark matter,
whereas $\Gamma < 0$ implies the reverse. This transfer of energy 
results in the effective equation of state of DE being phantom-like ($w_{\rm eff} < -1$) for 
$\Gamma < 0$ and quintessence-like ($w_{\rm eff} > -1$) for $\Gamma > 0$.
Note that in both cases DE has the EOS of the cosmological constant,
namely $w=-1$. However, 
the fact that the density of $\Lambda$ is growing/decreasing at the expense of that of matter
leads to either $w_{\rm eff} < -1$ when $\rho_{\rm DM} \to \rho_\Lambda$ ($\Gamma < 0$)
 or to $w_{\rm eff} > -1$ when $\rho_\Lambda \to \rho_{\rm DM}$ ($\Gamma > 0$).

\begin{figure*}
\begin{center} 
\resizebox{220pt}{150pt}{\includegraphics{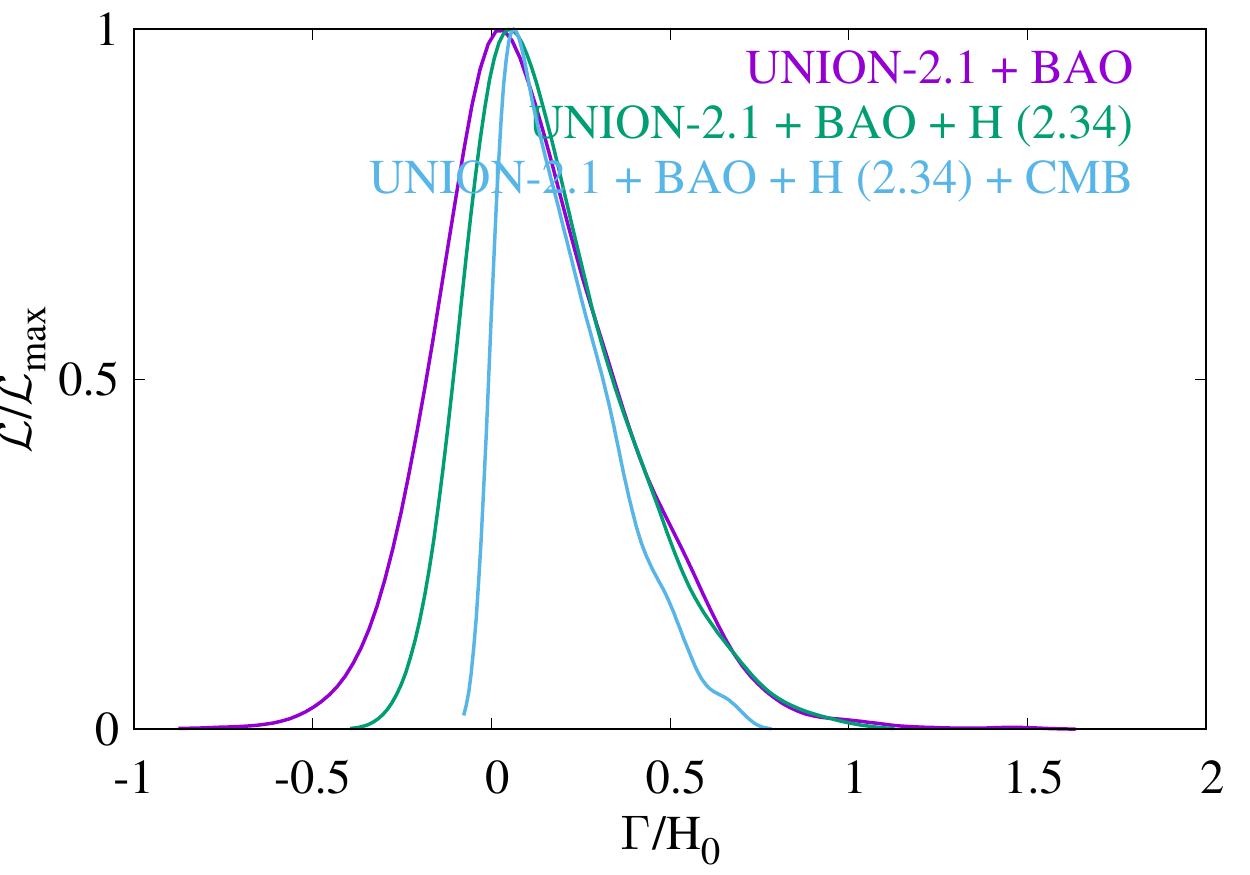}}\hskip -5 pt 
\resizebox{220pt}{150pt}{\includegraphics{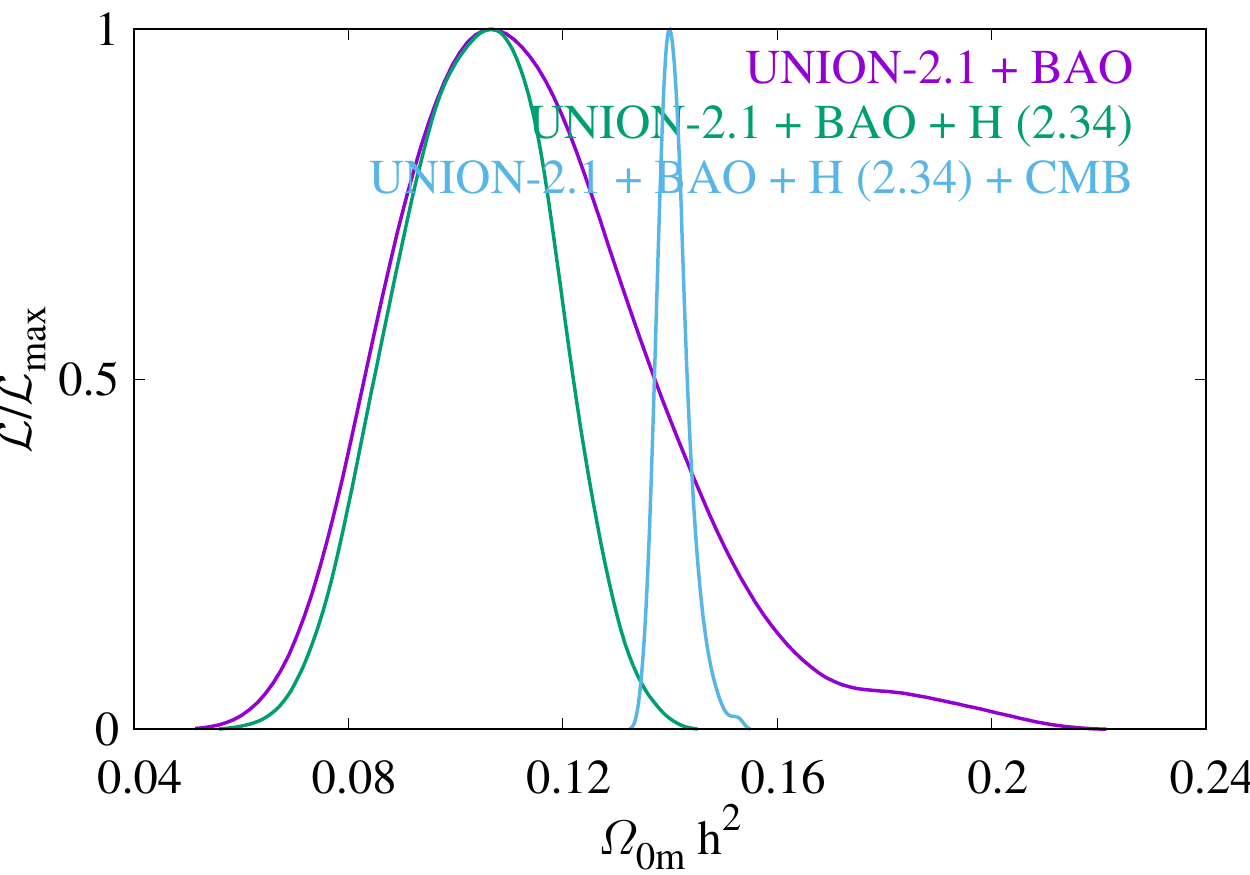}} 

\resizebox{220pt}{180pt}{\includegraphics{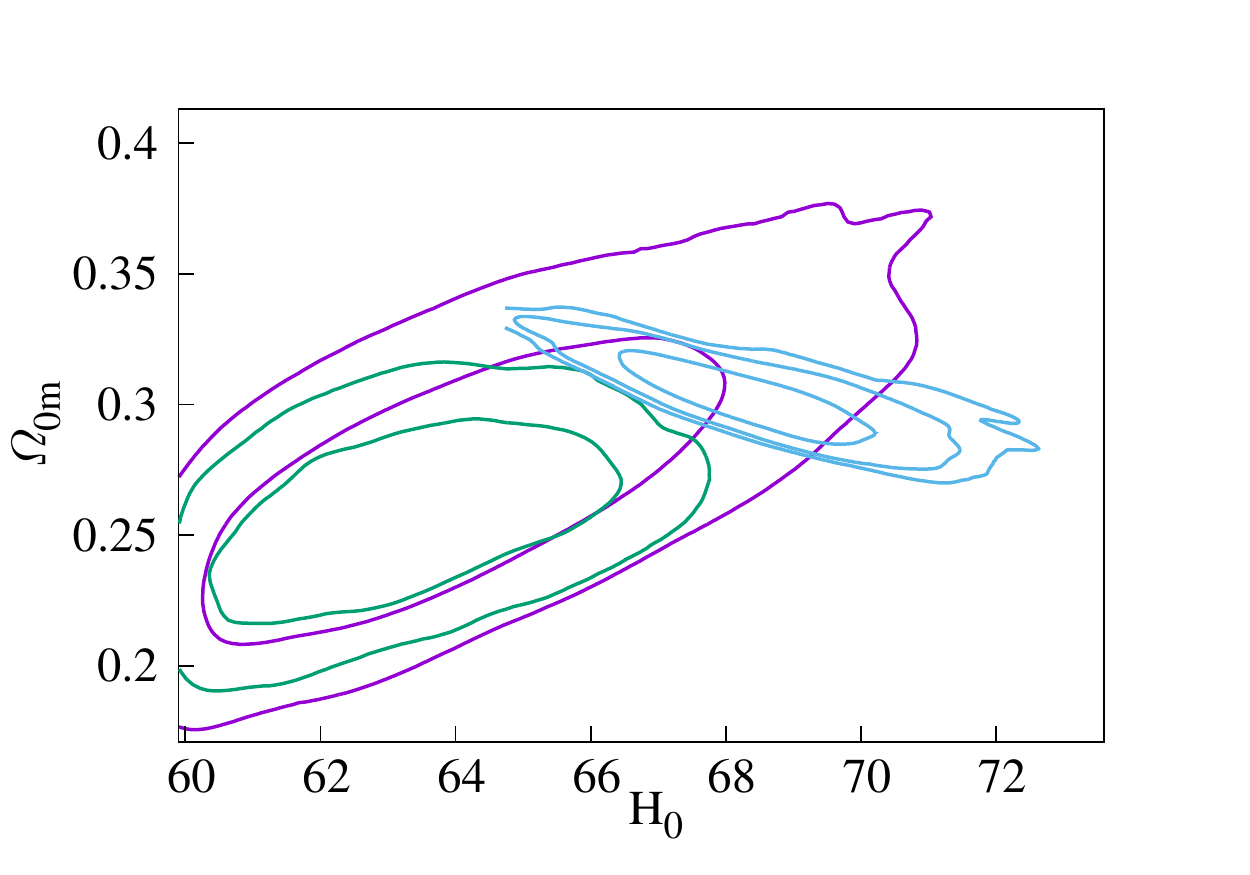}} \hskip -5 pt
\resizebox{220pt}{180pt}{\includegraphics{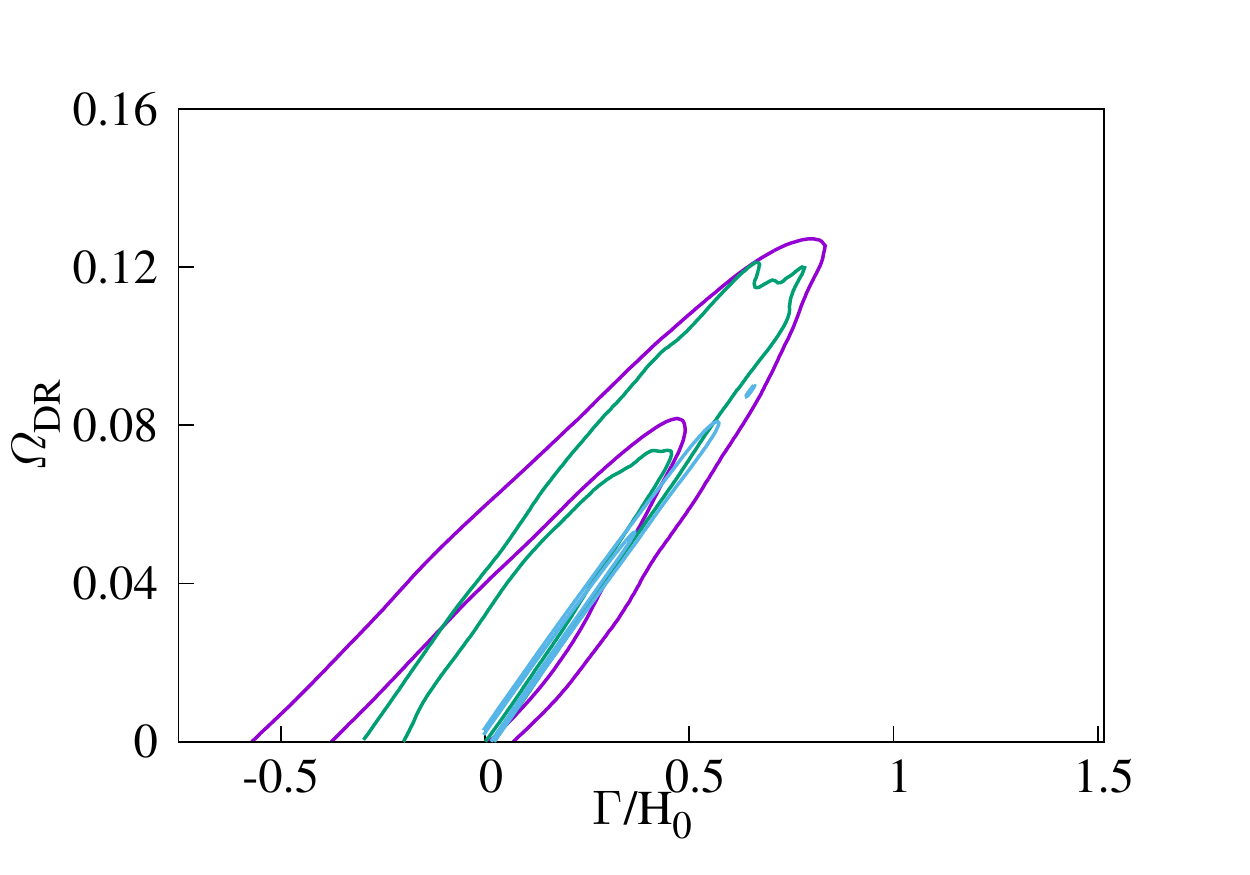}}

\end{center}
\caption{\label{fig:cntrs-model-III} Results for Model III. 
[Top left] One dimensional marginalized likelihoods of the decay parameter $\Gamma$ for model III (Eq.~\ref{eq:model-III}). 
[Top right] 1D likelihoods of $\Omega_{\rm 0m}h^2$.
[Bottom left] 2D contours of $\lbrace\Omega_{\rm 0m},H_0\rbrace$.
[Bottom right] 2D contours of $\lbrace\Omega_{\rm DR},\Gamma/H_0\rbrace$. Model III seems to fail in alleviating
 the tension between the datasets since radiation density is tightly constrained. 
 %by the high redshift observations (CMB). 
%{\bf Varun: the last sentence is unclear to me. This model is producing radiation at low redshift when
%$\Gamma < 0$, so why
%should high-$z$ measurements constrain it tightly, unless $\Gamma > 0$ is explicitely assumed ?}
} 
\end{figure*}

\begin{figure*}
\begin{center} 

\resizebox{145pt}{120pt}{\includegraphics{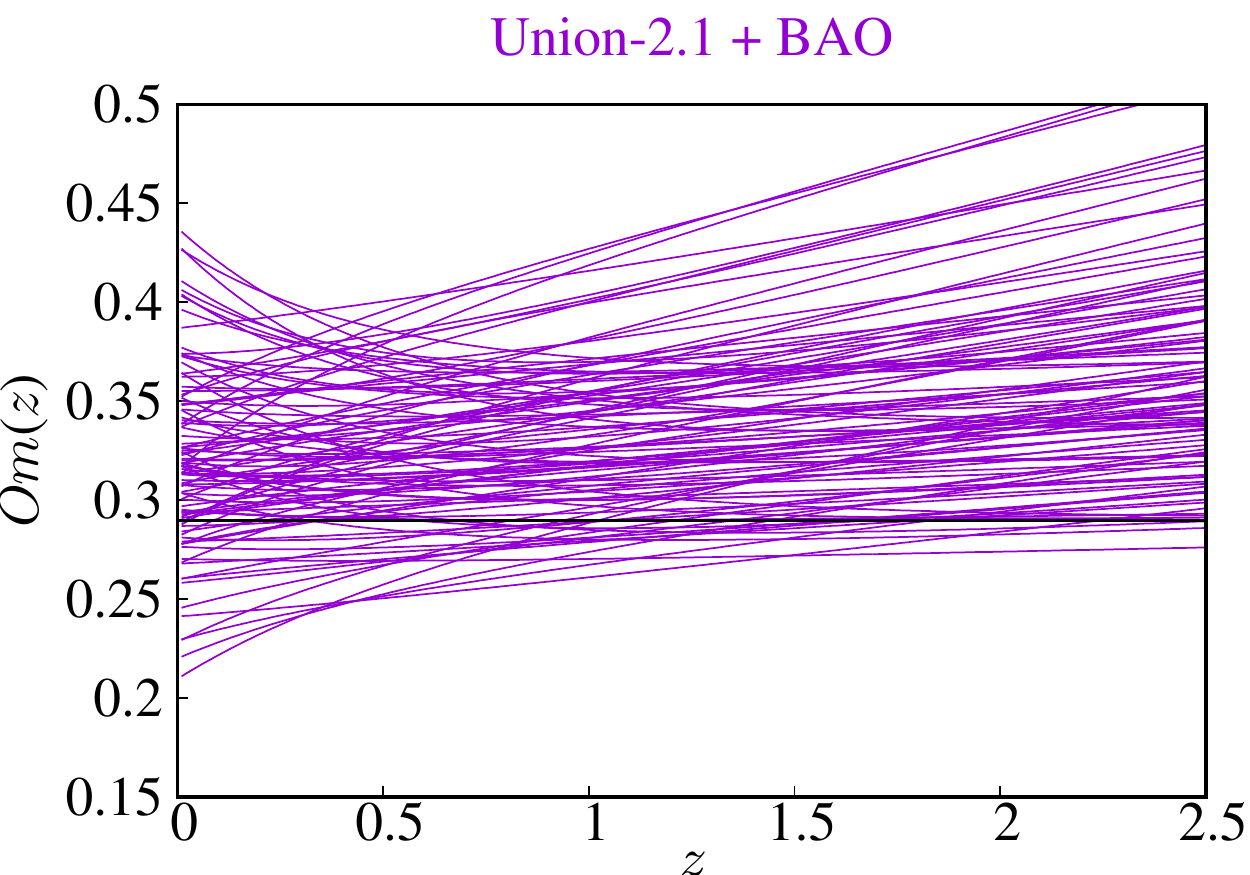}} 
\hskip -5pt \resizebox{145pt}{120pt}{\includegraphics{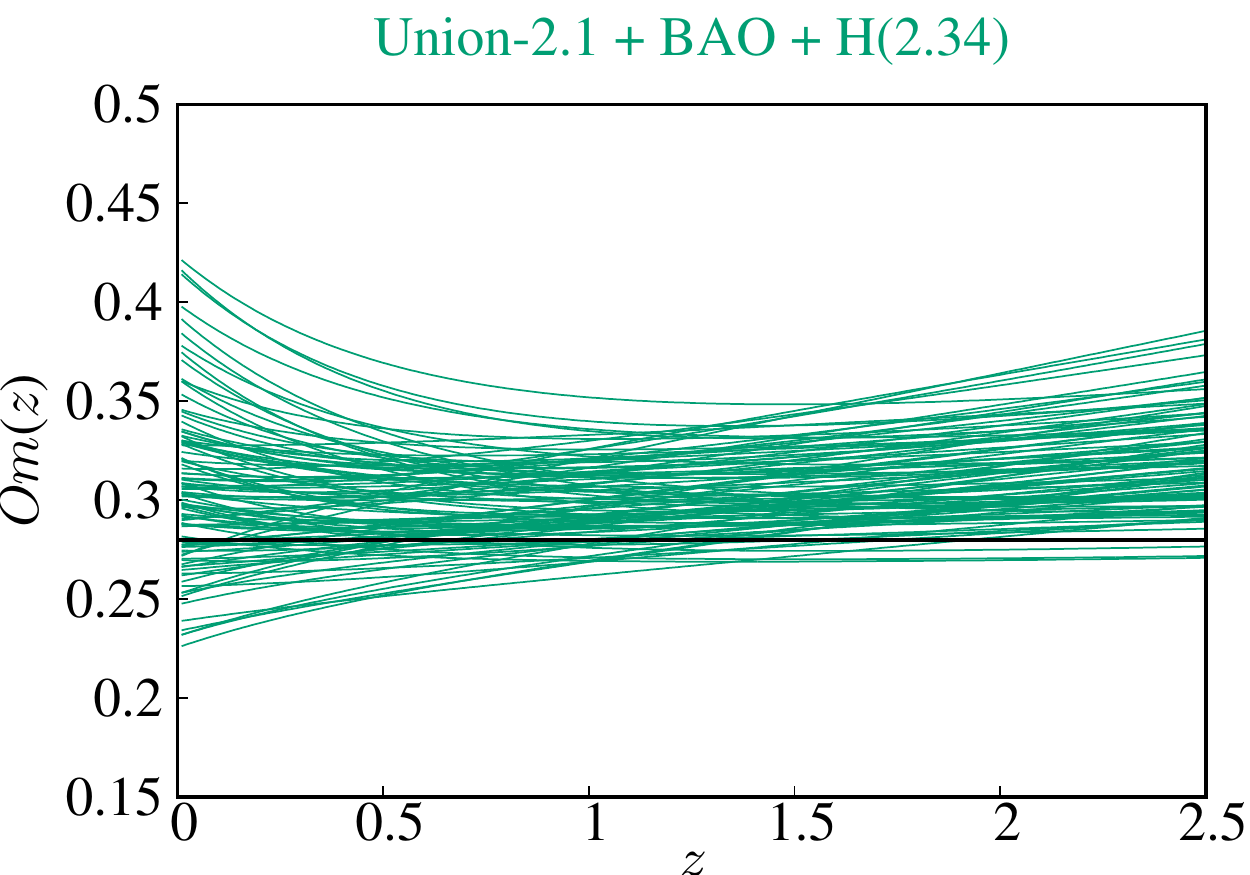}} 
\hskip -5pt \resizebox{145pt}{120pt}{\includegraphics{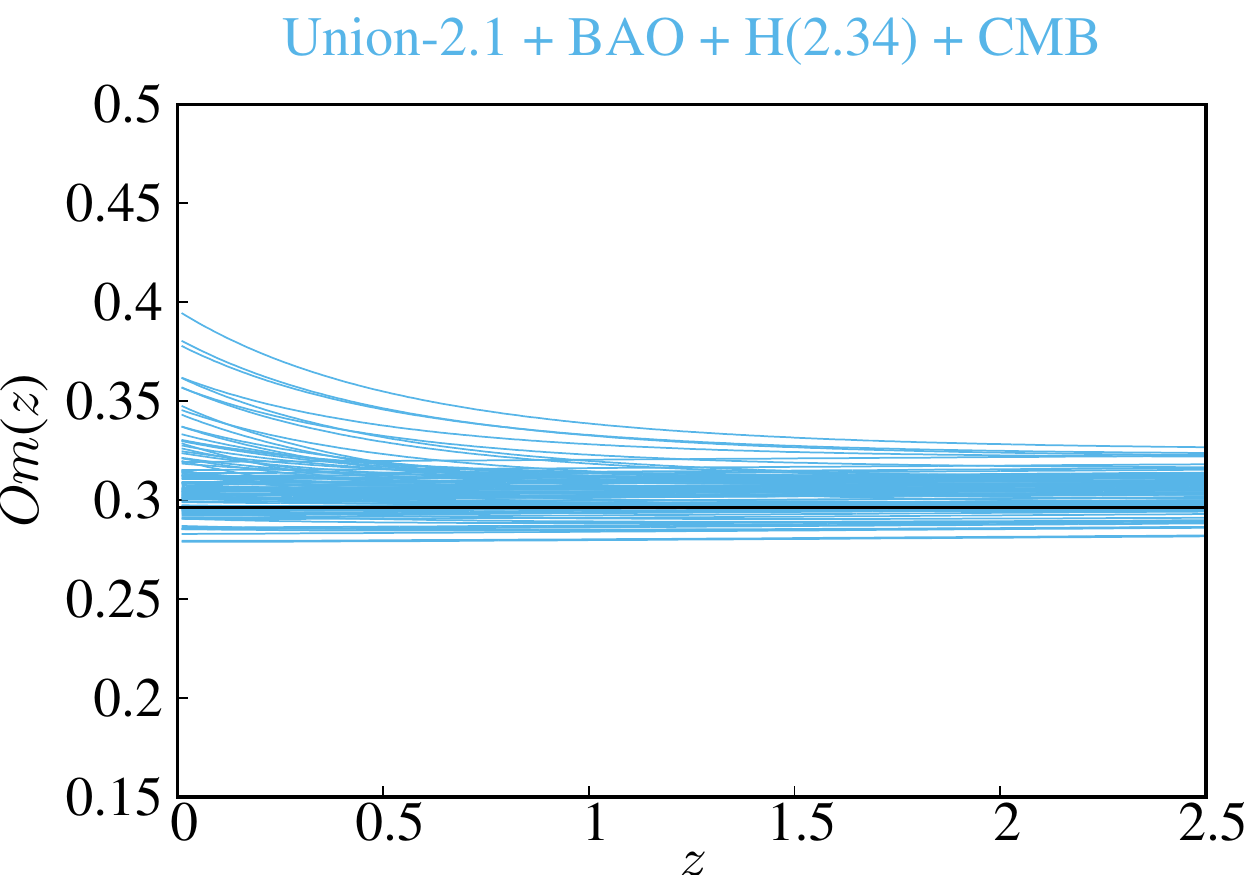}} 

\resizebox{145pt}{120pt}{\includegraphics{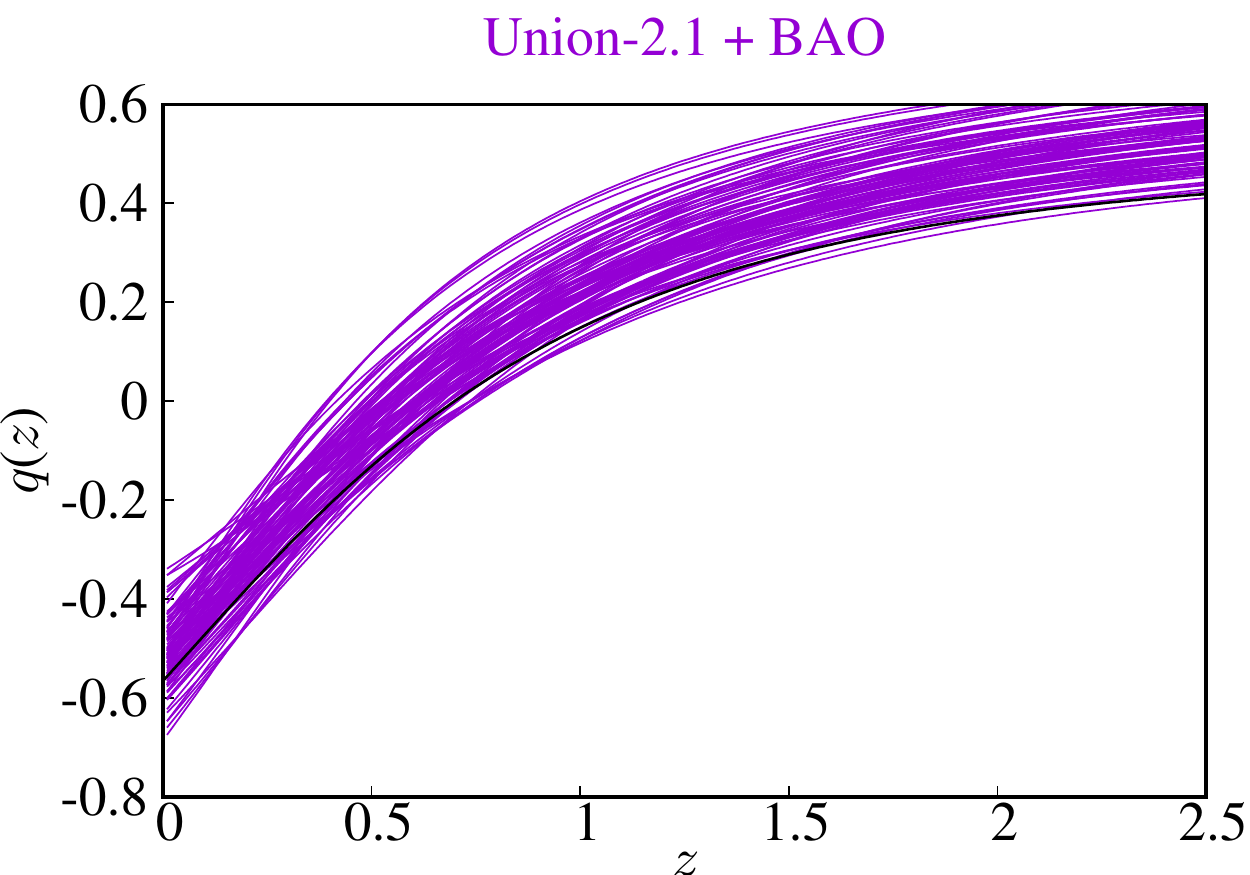}} 
\hskip -5pt \resizebox{145pt}{120pt}{\includegraphics{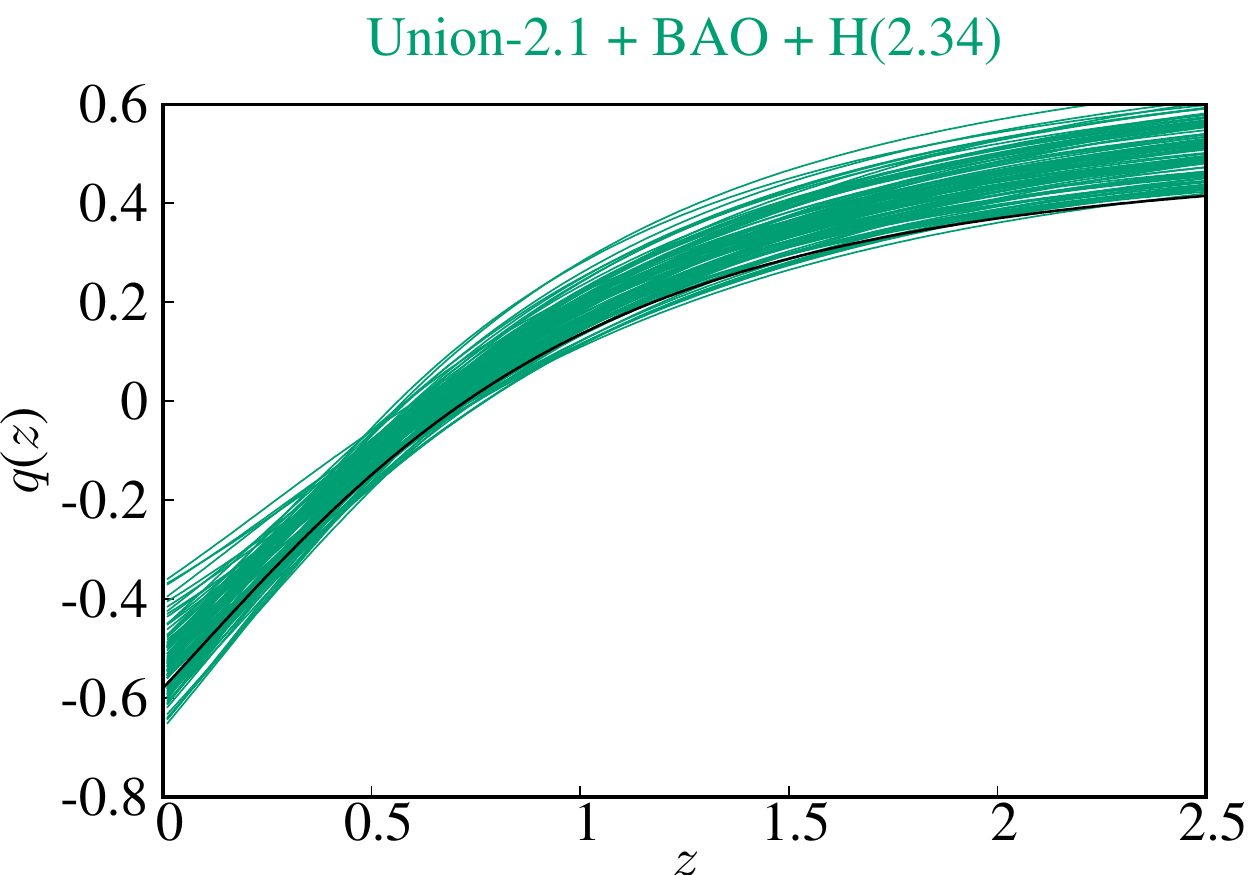}} 
\hskip -5pt \resizebox{145pt}{120pt}{\includegraphics{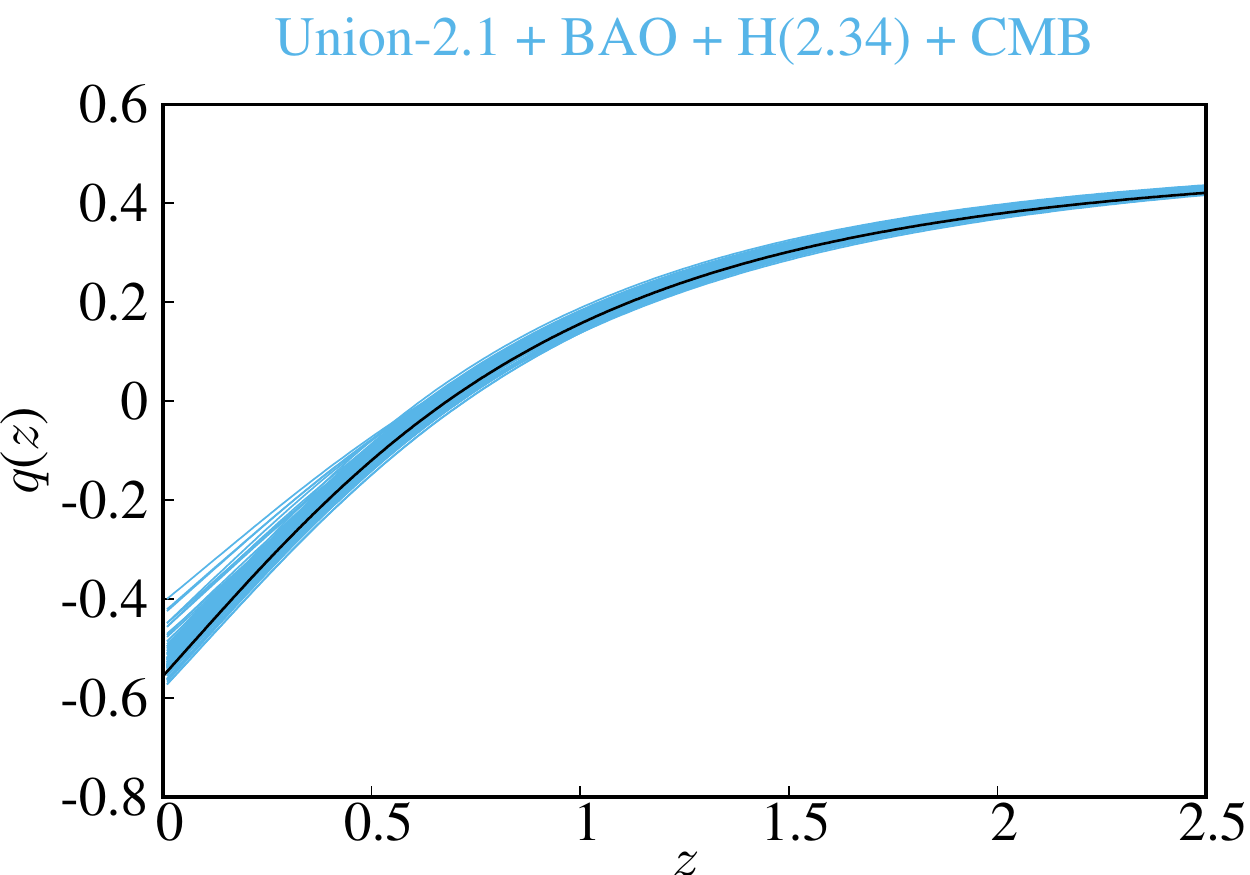}} 

\end{center}
\caption{\label{fig:sample-model-III} Samples $Om(z)$ and $q(z)$ as a function of redshift for model-III (Eq.~\ref{eq:model-III}). The datasets are 
indicated at the top of each plot. The black lines correspond to the best fit $\Lambda$CDM, for the same combination
of datasets.}
\end{figure*}

Figures~\ref{fig:cntrs-model-III} and \ref{fig:sample-model-III}  show results for Model III (DE decaying into 
dark radiation). This model has a particular characteristic which precludes large flexibility
 if one considers high redshift data (such as CMB). In fact observational constraints set stringent limits on the
 amount of radiation density in the past. 
This is due to the fact that
 the radiation density increases by $(1+z)^4$, consequently its inferred value
at high redshifts can change considerably if we change its current density by pumping energy
from it into DE or vice versa. Confronted with the data, Model III show a tiny possibility that dark energy might decay to dark radiation in small amounts. 
Our results also show that in this model one cannot alleviate the tension between the 
$H(2.34)$ data point and CMB data. So in this sense model III resembles $\Lambda$CDM. This model also posed us with some technical difficulties to derive the expansion history for different points in its parameter space due to divergence of some quantities. 

In table~\ref{tab:chi2} we show the derived values of the cosmological parameters for all three models. 
It is interesting that for Model I and II the
 data permits a greater flexibility in the selection of cosmological parameters
than $\Lambda$CDM.
Also in these models, and especially in Model II, the tension between CMB and 
the $H(2.34)$ data point is absent.

\begin{table*}
\begin{scriptsize}
\begin{center}
\tabcolsep=0.05cm
  \hspace*{0.0cm}\begin{tabular}
  {c|c|c|c } \hline 
  
  Data & Union-2.1 + BAO & Union-2.1 + BAO &Union-2.1 + BAO   \\ 
    & & + H (2.34)  &+ H (2.34) + CMB  \\ \hline

  %%%%%%%%%%%%%%%%%%%%%%%%%%%%%%%%%%%%%%%%%%%%%%%%%%%%%%%%%%%%%%%%%%%%%%%%%%%%%%%%%%%%%%%%%%%%%%%%%%%%%%%%%%%%%%%%%%%
%   ${\Omega_{\rm b}h^2}$ 	& 0.0216& 0.0216&  0.0222   \\  
% 				&$0.0216\pm0.0012$  & $0.0216\pm0.0012$&$0.0222\pm0.00024$\\  \hline

  ${\Omega_{\rm 0CDM}h^2}$ 	& 0.1071&0.1017 & 0.1166  \\  

				&$0.1119\pm0.026$  & $0.1023^{+0.0075}_{-0.0077}$&$0.1166\pm0.0017$\\  \hline

  ${\Omega_{\rm 0m}h^2}$ 	& 0.129 & 0.124 &0.139\\  
				&$0.134^{+0.02}_{-0.026}$  &$0.124^{+0.008}_{-0.007}$ &$0.1394\pm0.0016$\\  \hline

  {$\Omega_{\rm 0m}$}		&0.29& 0.28&0.296\\  
				&$0.294^{+0.033}_{-0.037}$       &$0.282^{+0.015}_{-0.017}$&$0.297\pm0.01$  \\  \hline

  {$H_0$}			&66.78 & 66.48& 68.59 \\  
				&  $67.27^{+2.13}_{-2.6}$&$66.48\pm1.3$ &$68.6^{+0.77}_{-0.8}$ \\ \hline 

  {$-2~\ln {\cal L}_{\rm max}$}			&546.1 &546.2 &550.25  \\ \hline 

  \end{tabular}
  \end{center}
\caption{~\label{tab:chi2lcdm}Best fit $\chi_{\rm best~fit}^2$ and cosmological parameters obtained for different datasets for $\Lambda$CDM model. The best fit values (first row in each parameter) and the mean with 1$\sigma$ 
deviations (second row in each parameter) are also provided for some parameters.
%{\bf Definition of $\Omega_{\rm 0m}$ in terms of $\Omega_{\rm 0b}$ and $\Omega_{\rm CDM}$ needs to be made clear. }
}
\end{scriptsize}
\end{table*}

\begin{table*}
\begin{tiny}
\begin{center}
\tabcolsep=0.04cm
  \hspace*{0.1cm}\begin{tabular}
  {c|c c c|c c c|c c c } \hline 
  
  Data &  & Union-2.1 &  & &Union-2.1 + BAO  & & &Union-2.1 + BAO & \\ 
   &  & + BAO&  & & + H (2.34) & & &  + H (2.34) + CMB & \\ 
  \cline{2-10}\\
  & Model I & Model II & Model III &Model I & Model II & Model III&Model I & Model II & Model III \\ \hline

  %%%%%%%%%%%%%%%%%%%%%%%%%%%%%%%%%%%%%%%%%%%%%%%%%%%%%%%%%%%%%%%%%%%%%%%%%%%%%%%%%%%%%%%%%%%%%%%%%%%%%%%%%%%%%%%%%%%
%   ${\Omega_{\rm b}h^2}$ 	&0.0215  &0.0215  &0.0217  
% 							& 0.0216&0.0214 & 0.0218
% 							&0.0222 & 0.0223&   0.0222 \\  
% 				&$0.0215\pm0.0012$  &$0.0216\pm0.0012$  &$0.0218\pm0.0012$ 
% 				& $0.0216\pm0.0012$&  $0.0215\pm0.0012$& $0.0218\pm0.0012$
% 				& $0.0222\pm0.00028$&$0.0223\pm0.0003$ &$0.0223\pm0.00025$    \\  \hline

  ${\Omega_{\rm 0CDM}h^2}$ 	&0.111  & 0.113 &  0.104
								& 0.102&0.116 &0.1
								&0.116 & 0.116&  0.117\\  

				&$0.119^{+0.027}_{-0.038}$  &$0.115^{+0.028}_{-0.035}$ & $0.093^{+0.018}_{-0.028}$
				&$0.103\pm0.008$&$0.109^{+0.022}_{-0.03}$& $0.081^{+0.015}_{-0.014}$
				& $0.116\pm0.002$&   $0.116\pm0.002$ &$0.119^{+0.002}_{-0.004}$ \\  \hline

  ${\Omega_{\rm 0m}h^2}$ 	&0.133  & 0.135& 0.126
							& 0.124&0.138 & 0.122
							&0.139 & 0.138&  0.139\\  
				&$0.141^{+0.027}_{-0.038}$  & $0.137^{+0.028}_{-0.035}$&$0.114^{+0.018}_{-0.028}$ 
				&$0.123^{+0.01}_{-0.007}$&$0.13^{+0.022}_{-0.03}$ & $0.103^{+0.015}_{-0.014}$
				& $0.139\pm0.002$& $0.138\pm0.002$&$0.14^{+0.002}_{-0.004}$  \\  \hline
				
{$\Omega_{\rm dr}h^2$}		&-  & -& 0 
							&- & -&0.01 
							&- & -&   0 \\  
				&-   &- &$0.016^{+0.018}_{<} $ 
				&- &- &$0.017^{+0.006}_{<}$ 
				&-&-&  $0.013^{+0.005}_{<}$  \\  \hline

  {$\Omega_{\rm 0m}$}		&0.292  & 0.296& 0.284 
							& 0.282& 0.298& 0.282
							&0.298 & 0.297& 0.298   \\  
				&$0.3^{+0.038}_{-0.04}$ &$0.298\pm0.04$ &  $0.28^{+0.03}_{-0.04}$   
				&$0.283^{+0.015}_{-0.017}$ &$0.28\pm0.03$  &$0.26^{+0.028}_{-0.024}$ 
				&$0.297\pm0.013$ &$0.299\pm0.01$ &$0.3\pm0.01$    \\  \hline

{$\Omega_{\rm dr}$}		&-  & -& 0 
							&- & -& 0.02
							&- & -&   0\\  
				&-   &- &   $0.04^{+0.01}_{<}$  
				&- &- &$0.04^{+0.01}_{<}$ 
				&-&-&  $0.03^{+0.008}_{<}$  \\  \hline

  {$H_0$}			&67.52  & 67.6 & 66.8
					& 66.26&68 &66 
					&68.34&68.2 &   68.5 \\  
				&$68.12^{+3.9}_{-4.5}$ & $67.5\pm3.4$ &$64.4^{+2}_{-3.1}$ 
				& $66.4\pm1.9$&$67^{+2.8}_{-3.5}$  & $63.4^{+1.75}_{-2.04}$
				&$68.5\pm1.6$ &$68.1^{+1}_{-1.1}$ &$68.5\pm1.2$    \\ \hline 

  {$\Gamma/H_0$}		& -0.085 &0.13 & -0.12 
						&0.03&0.16 &0.13
						&0.046& -0.011&0.005    \\  
				&$-0.1^{+0.45}_{-0.38}$ & $0^{+0.39}_{-0.28}$&  $0.12^{+0.2}_{-0.3}$
				& $0.028\pm0.3$& $0.04^{+0.34}_{-0.3}$&$0.17^{+0.14}_{-0.27}$ 
				&$0.02\pm0.25$ &$-0.015\pm0.02$&$0.2^{+0.05}_{-0.19}$    \\  \hline\hline
{$-2~\ln {\cal L}_{\rm max}$}	& 546.1  & 546& 546.1
		&546.2 &546 &546.2 
		&550.2&550&  550.2\\  \hline 
  
  \end{tabular}
  \end{center}
\caption{~\label{tab:chi2}Best fit $\chi_{\rm best~fit}^2$ obtained for different
models considering different combination of datasets. The best fit values of cosmological parameters is shown in the 
first row, and the mean (with 1$\sigma$ deviations) is shown in the second row. 
P.C. refers to the situation where the posterior distribution is cut in the prior range. A `$<$' symbol denotes that the 
parameter is unbounded from below. 
%{\bf Varun: what is the significance of the last row called `Total', 
%is this the value of $-2~\ln {\cal L}_{\rm max}$ ? Also what does $<$ for Model III (for $\Omega_{de}$)
%signify ?} {\color{red} I think we will not show all these numbers in the draft. In fact none of the models can fit the data significantly better. However, we can focus on the agenda of this work to discuss a new calss of dark energy models that can explain observations.}
}
\end{tiny}
\end{table*} 

\section{Discussion}\label{sec:discussion}

In this paper we propose a new class of  dark energy models in which dark energy can decay, either exponentially in time, 
or into dark matter or into dark radiation with a time-independent rate.
One might note that 
decaying DE models have been earlier discussed in the context of 
dark matter-dark energy interactions~\citep{Amendola:1999er,Guo:2007zk,Boehmer:2008av,Valiviita:2008iv,He:2008tn,Micheletti:2009pk,He:2010im,Pavan:2011xn,Faraoni:2014vra,Salvatelli:2014zta,Abdalla:2014cla},
modified gravity~\citep{Sahni:2002dx,Shtanov:2009ss}
and quintessence~\citep{Gu:2001wp,Alam:2003rw,Kallosh:2003bq,Blais:2004vt,Wang:2004nm,Dutta:2008qn,Gupta:2011ip,Bolotin:2012yja}.
The specialty of our model is that the decay of DE is related to the intrinsic properties
of DE
and not to the expansion of the Universe. 
Thus the properties of DE do not depend upon cosmological expansion 
or on the presence of a specific form of
the DE potential or even on the equations governing cosmological expansion 
(viz. FRW, modified gravity, etc).

In a certain sense our class of models have much in common with the
radioactive decay of matter. In similar fashion, one can describe the decay of DE in terms of a 
fundamental constant,
$\Gamma$, which is related to the `half-life' of DE as $t_{1/2} = \ln(2)/\Gamma$.
 The three models which we consider, though similar in equations and mechanism,
give rise to somewhat different observational predictions. Model I describes dark energy 
decaying exponentially and as a result it has an evolving effective equation of state. 
Such models can be viable for a large region in parameter space and can {\em marginally} 
alleviate the tension between CMB and QSO based $H(2.34)$ BAO data which is faced by $\Lambda$CDM. 

\begin{figure*}
\begin{center} 

\resizebox{165pt}{120pt}{\includegraphics{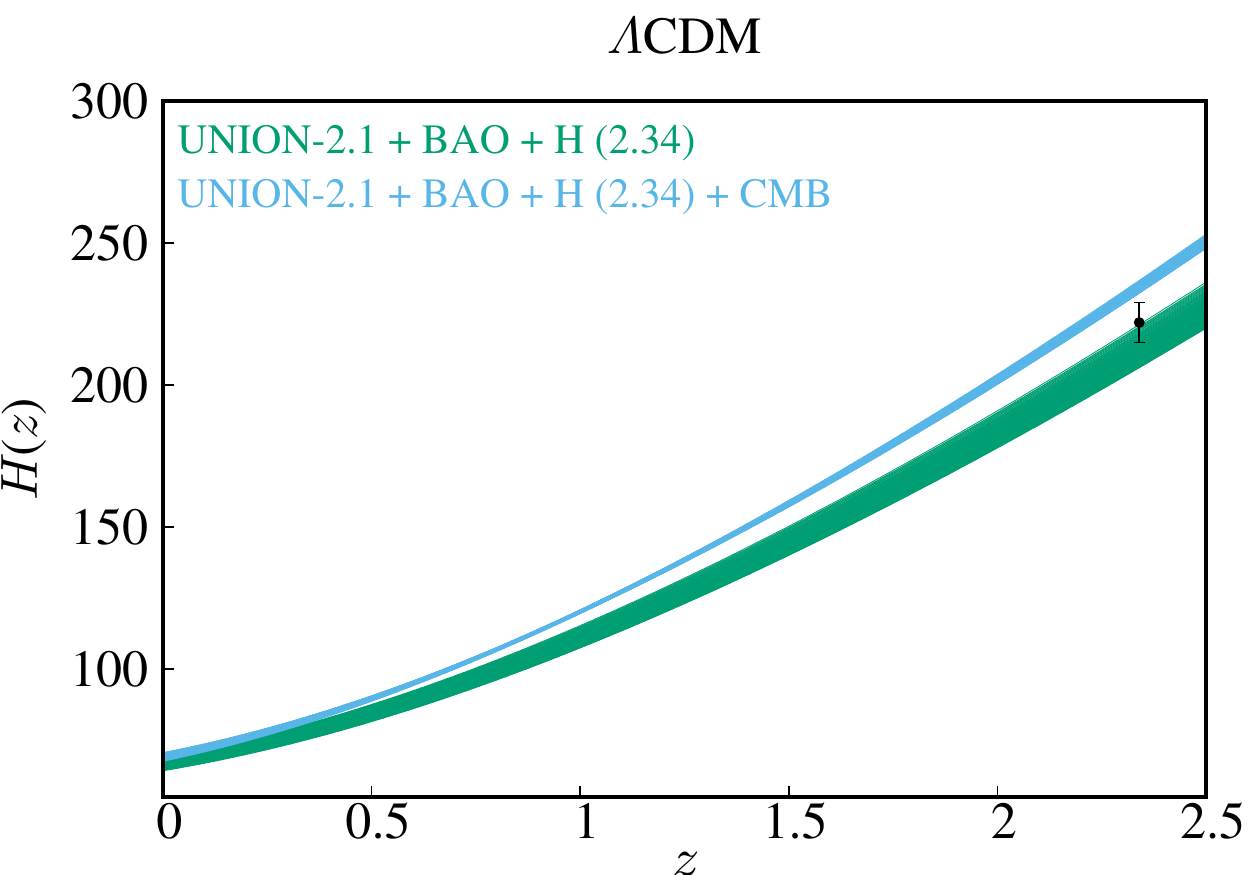}} 
\resizebox{165pt}{120pt}{\includegraphics{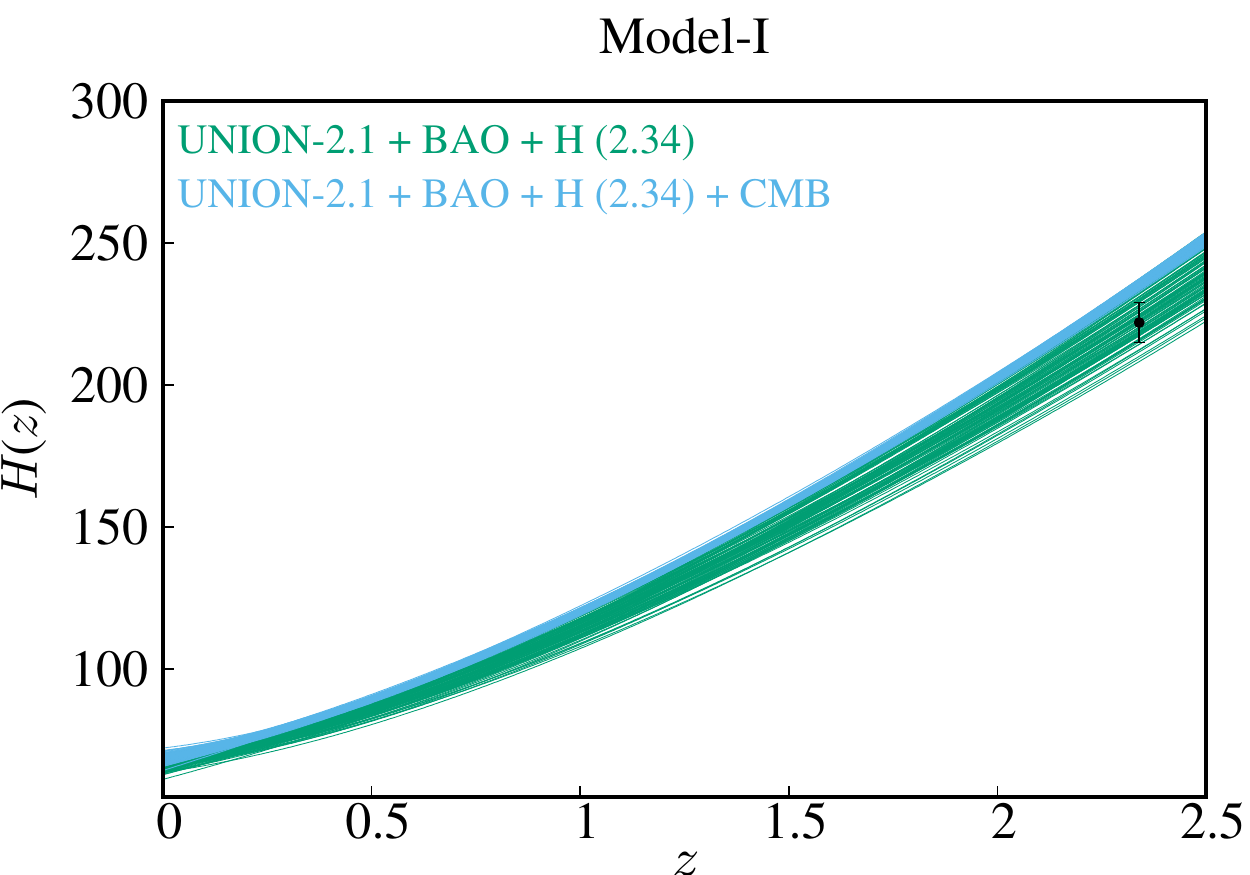}} 

\resizebox{165pt}{120pt}{\includegraphics{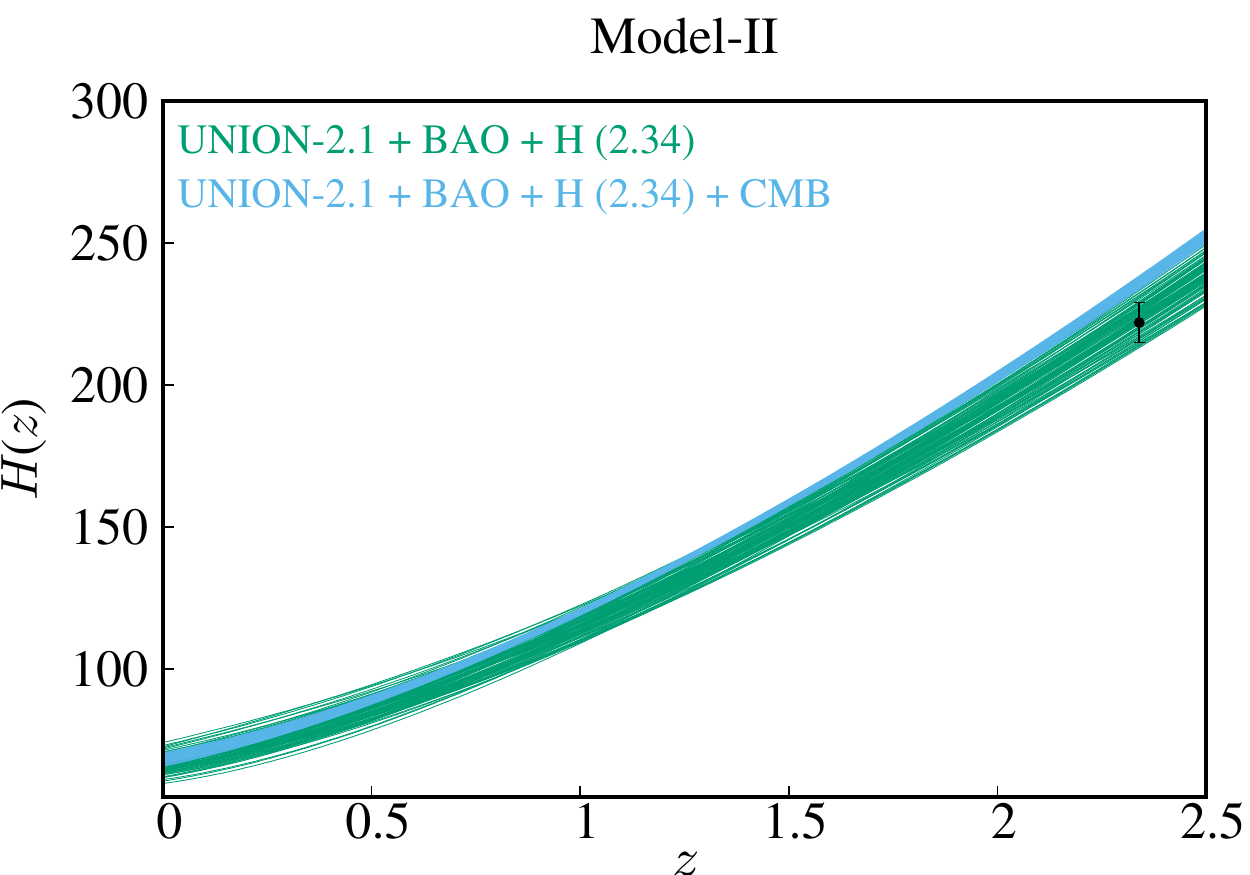}} 
\resizebox{165pt}{120pt}{\includegraphics{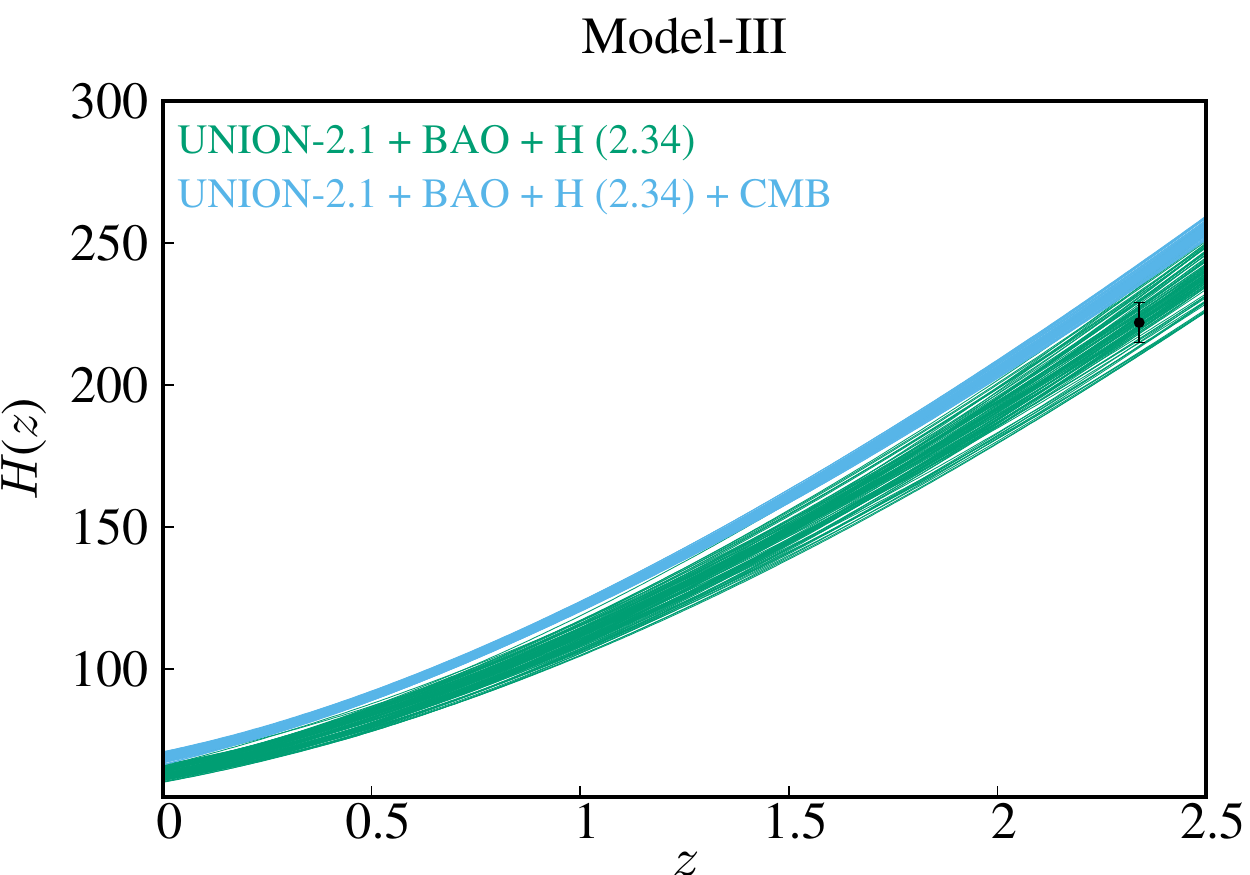}} 
\end{center}
\caption{\label{fig:HZ-samples} Samples $H(z)$ as a function of redshift for different models are provided for 2 different dataset combination that include $H(z=2.34)$
measurement. The datasets are indicated at the top of each plot. Note that we have 100 samples in each case representing the allowed region within 2$\sigma$. We find that
for model-I and model-II there are large overlap between the samples while for $\Lambda$CDM and model-III, the samples show discordance, especially in the former case. 
The QSO BAO data is plotted in black with errorbar.}
\end{figure*}
 
 %Since both signs of $\Gamma$ 
%are allowed it follows that the decay of $\Lambda$-like dark energy into dark energy with 
%a stiffer equation of state is permissible, but so is the opposite
% possibility of dark energy which emerges at low redshifts.
%{\bf Varun: As I mentioned earlier, the discussion on model I is incomplete since we do not talk
%about its decay products.}
%{\color{red} I agree that we should have some more discussion to elaborate better the case of Model I in the paper.}

 In Model II, dark energy decays into dark matter. In this model the axis of confidence contours (derived from confronting the model with different sets of cosmological data) 
is rotated with respect to the case of Model I. %This interesting phenomenon is due to the different natures of these two models. 
Interestingly, data analysis using Model II shows absolutely no tension between any two sets of cosmology data.
Indeed, the derived confidence contours all show proper overlap with each other -- see
figure~\ref{fig:cntrs-model-II}. Incorporating CMB data into the analysis we
find that Model II peaks strongly around the cosmological constant, that sets strong limits on the half-life of dark energy, namely
that it needs to be several times the age of the Universe. 

In Model III dark energy decays into dark radiation (DR). 
Cosmological observations place tight constraints on this model,
especially when $\Gamma < 0$ which corresponds to DR decaying to DE and implies 
an increase in the radiation density at high redshift. 
%Our results for $\Gamma < 0$ show that still some decay of dark radiation 
%into dark energy at low redshifts is allowed which  could lead to the slowing down of cosmic acceleration. In fact all three models permit the possibility the cosmic acceleration slowing down 
%at low redshifts, which was first pointed out in \citep{arman2009};  for a more recent discussion see \citep{slow_down2015} and references therein. 

%{\bf Varun:  I don't think we have shown any evidence of `cosmic acceleration slowing down'
%in any of our figures. So perhaps the last sentence is redundant.}
In figure~\ref{fig:HZ-samples} we plot the $H(z)$ samples within the 2$\sigma$ confidence level for different models and for different dataset combination that includes
the QSO BAO data. Top left plot show the $\Lambda$CDM model results where we find the samples for the dataset combination with and without CMB data have no overlap. Top right plot and bottom 
left plot show $H(z)$ for model-I and model-II respectively. Note that both these models have substantial overlap of samples for Union-2.1 + BAO + H(2.34) and Union-2.1 + BAO + H(2.34) + CMB
dataset combinations, suggesting no particular tension between CMB and QSO data. However as has been pointed out earlier, model-III does not help in alleviating the tension and the samples 
do not show consistency (unlike model-I and II).

 To summarize, model II -- in which DE decays to dark matter, is perhaps the most
compelling of the three models which have been studied. All the three data sets which
we consider, namely SNIa, BAO and CMB show consistency with each other for this
fiducial cosmology and the $2\sigma$ upper limit on $|\Gamma|/H_0$ is less than $0.041$ if all
data are taken into account. Thus, since $H_0t_0 \sim 1$ for the best-fit $\Lambda$CDM model, this
means that the half-life time for this channel of DE decay $t_{1/2} >
17 t_0\approx 7\times 10^{18}$ s where $t_0$ is the present age of the
Universe! For the other two models, some tension exists, especially between the
CMB and the Lyman-$\alpha$ derived BAO at $z=2.34$. Due to this tension, the obtained $2\sigma$ limits on $|\Gamma|/H_0$ for 
these channels of decay are not so strong as in the previous case. Still
they both are significantly less than unity.

\section*{Acknowledgments}
A.S. would like to acknowledge the support of the National Research Foundation of Korea (NRF-2016R1C1B2016478). 
DKH acknowledges Laboratoire APC-PCCP, Universit\'e Paris Diderot and Sorbonne Paris Cit\'e (DXCACHEXGS) and 
also the financial support of the UnivEarthS Labex program at 
Sorbonne Paris Cit\'e (ANR-10-LABX-0023 and ANR-11-IDEX-0005-02). A.A.S. was partially supported by the grant RFBR 17-02-01008 and by the 
Scientific Program P-7 of the Presidium of the Russian Academy of Sciences.\\
%D.K.H and A.S wish to acknowledge support from the Korea Ministry of Education, Science
%and Technology, Gyeongsangbuk-Do and Pohang City for Independent Junior Research Groups at 
%the Asia Pacific Center for Theoretical Physics. We also acknowledge the use of publicly 
%available CAMB and CosmoMC in our analysis.
\bibliographystyle{mnras}
\bibliography{RDE}

\appendix
\section{$h^2(z)$ in the large decay time limit}

Though to determine the evolution law $h^2(z)$ we have solved Eqs. (2.3), (2.4-2.6) and (2.7-2.9) exactly, to understand the answers it is instructive to consider 
the limit of large decay times $|\Gamma|\ll H_0$ which is both expected and follows from our calculations. Then it becomes possible to present explicit expressions for $h^2(z)$.

 \subsection{Model I}   

For $\Gamma t \ll 1$, we get in the first order:

\begin{eqnarray}
\rho_{DE}=\epsilon_0e^{-\Gamma t}\approx \epsilon_0(1-\Gamma t) \\
h^2(z) = \Omega_{0m}(1+z)^3 + (1-\Omega_{0m})[1-\Gamma (t-t_0)]
\end{eqnarray}

Here the $t(z)$ dependence in the last term can taken from the $\Lambda$CDM model with the same value of $\Omega_{0m}$:

\begin{equation}
t(z)=t_0-H_0^{-1}\int_0^z\frac{dz_1}{1+z_1}\left[1-\Omega_{0m}+\Omega_{0m}(1+z_1)^3\right]^{-1/2}
\end{equation}

Thus, with the same accuracy:

\begin{widetext}
\begin{equation}
\hspace{50 mm} h^2 (z)= 1-\Omega_{0m}+\Omega_{0m}(1+z)^3 + \frac{\Gamma(1-\Omega_{0m})}{H_0}\int_0^z\frac{dz_1}{1+z_1} \left[1-\Omega_{0m}+\Omega_{0m}(1+z_1)^3\right]^{-1/2}
\label{h2Model1}
\end{equation}
\end{widetext}
It is clearly seen that $h^2(z)$ in this model is always {\em larger} in the past compared to the standard $\Lambda$CDM model with the same value of $\Omega_{0m}$.

Let us now compare this result with what happens in the case of DE being slowly rolling quintessence, i.e. a scalar field $\phi$ with a potential $V(\phi)$ minimally coupled to gravity. The assumption $\Gamma t_0\ll 1 $ means that the change in $V(\phi)$ is small up to the present time $t_0$. So, we can expand $V(\phi)=V_0+V_{\phi 0}\psi$ where $\psi= \phi - \phi_0,~\phi_0=\phi(t_0),~V_0=V(\phi_0),~V_{\phi 0} = \frac{dV}{d\phi}(\phi_0)$. Then
\begin{widetext}
\begin{eqnarray}
\psi = V_{\phi}\int_{t}^{t_0}\frac {dt_1}{a^3(t_1)}\int_0^{t_1} a^3(t_2)\, dt_2,~~~\rho_{DE}= V_0+V_{\phi 0}\psi + \frac{\dot\psi^2}{2} \nn\\
h^2=\Omega_m(1+z)^3 + (1-\Omega_m)\left[1+\frac{V_{\phi 0}^2}{V_0}\int_{t}^{t_0}
\frac {dt_1}{a^3(t_1)}\int_0^{t_1} a^3(t_2)\, dt_2 +\frac{V_{\phi 0}^2}{2a^6V_0}\left(\int_0^t a^3(t_1)dt_1\right)^2 \right]
\label{h2Quint}
\end{eqnarray}
\end{widetext}

Once more, the $t(z)$ dependence in all integrals is taken from the $\Lambda$CMD model with the same value of $\Omega_m$. Though Eq.~(\ref{h2Quint}) is slightly more complicated than Eq.~(\ref{h2Model1}), consideration of the future de Sitter stage with $H_{dS}=H_0\sqrt{1-\Omega_m}$ (that occurs as far as $\Gamma t\ll 1$) shows that here

\begin{equation}
\Gamma = \frac{V_{\phi 0}^2}{3H_{dS}V_0}
\label{Gamma}
\end{equation}
\vspace{5mm}

\clearpage

\subsection{Model II \& III}
In the same approximation:

\begin{widetext}
\begin{eqnarray}
\dot\rho_{DE}&=&-\Gamma\rho_{DE}\nn\\
\Gamma &=&{\rm const}\nn\\
\dot\rho_m+3H\rho_m&=&\Gamma\rho_{DE}\nn\\ 
\rho_{DE}&\approx& \epsilon_0(1-\Gamma t),~~\rho_m\approx\frac{\rm const}{a^3}+\Gamma\epsilon_0\int_0^ta^3\, dt\nn\\ 
h^2(z)&=&1-\Omega_{0m}+\Omega_{0m}(1+z)^3+\Gamma\left[t_0-t(z)-a^{-3}\int_t^{t_0}a^3(t_1)dt_1\right]\nonumber \\
&=&1-\Omega_{0m}+\Omega_{0m}(1+z)^3- \frac{\Gamma(1-\Omega_{0m})}{H_0}\int_0^z \frac{dz_1}{1+z_1}
\left[1-\Omega_{0m}+\Omega_{0m}(1+z_1)^3\right]^{-1/2}\left[\left(\frac{1+z}{1+z_1}\right)^3-1\right] 
\end{eqnarray}
\end{widetext}
Here, we find just the opposite, $h^2(z)$ is {\em smaller} in the past compared to the standard case. This crucial difference between the models is due to different equations of state of products of DE decay in them and because of their normalization to the present relative amount of  dark matter $\Omega_{0m}$. It is straightforward to generalize this result to the model III with DE decay into dark radiation. In this case the situation appears to be the same as for the model I:  $h^2(z)$ was relatively larger in the past.

\bsp	% typesetting comment
\label{lastpage}
\end{document}